\documentclass[aps,prd,twocolumn,superscriptaddress]{revtex4-2}
\usepackage{graphicx}
\usepackage{amsmath}
\usepackage{amssymb}
\usepackage{multirow}
\usepackage{booktabs}
\usepackage{dcolumn}
\usepackage{slashed}
\usepackage{mathtools}
\usepackage[colorlinks,
linkcolor  = blue,
urlcolor   = blue,
citecolor  = blue,
anchorcolor= blue]{hyperref}
\usepackage{float}
\usepackage{subcaption}
\usepackage{ragged2e}



\usepackage{xcolor}
\usepackage{colortbl}
\definecolor{lightgray}{RGB}{230,230,230}
\usepackage{bm}
\usepackage{placeins}
\usepackage{multirow} 
\usepackage{graphicx}  
\usepackage{stfloats}
\usepackage{pdfcomment}
\begin{document}
\title{Hot and dense medium effects on the \(B_s^*\) and \(B^*\) multiplets}
\author{K.~Azizi}
\thanks{Corresponding author}
\email{kazem.azizi@ut.ac.ir}
\affiliation{Department of Physics, University of Tehran, North Karegar Avenue, Tehran
14395-547, Iran}
\affiliation{Department of Physics, Faculty of Engineering and Natural Sciences, Dogus University,
  Dudullu-\"{U}mraniye, 34775 Istanbul, T\"{u}rkiye}

\author{N.~Er}
\affiliation{Department of Physics, Bolu Abant İzzet Baysal University,
  G\"{o}lk\"{o}y Campus, 14030 Bolu, T\"{u}rkiye}

\author{J.Y.~S\"{u}ng\"{u}}
\affiliation{Department of Physics, Kocaeli University,
  41001 Izmit, T\"{u}rkiye}

\date{\today}

\begin{abstract}
We present an extensive analysis of the in-medium masses and decay
constants of the $B_s^*(5415)$ and $B^*(5325)$ multiplets, including
both particles and antiparticles, using QCD sum rules at finite
temperature and density. The OPE incorporates the full temperature-
and density-dependent contributions from the quark, gluon, and mixed
condensates. Computing the strange ($B_s^{*0}$, $\bar{B}_s^{*0}$),
charged ($B^{*\pm}$), and neutral ($B^{*0}$, $\bar{B}^{*0}$) doublet
properties allows us to study the effects of flavor symmetry breaking,
strangeness, and heavy-quark decoupling on the beauty vector mesons in
the medium. Our results indicate that the mass is remarkably resistant
to the medium across the entire multiplet: no state loses more than
$\sim 13\%$ of its vacuum value, even at $T = T_c$ and $n = 5n_0$, the
extreme conditions explored here. The decay constant is far more
sensitive, losing up to $\sim 78\%$ at the same point. Baryon density
clearly dominates the medium response, while temperature plays a
secondary role until the system approaches the deconfinement crossover.
At zero density, every state loses almost the same fraction of its mass
and decay constant: mass shifts lie between $-(0.5$-$1.1)\%$ and
decay-constant shifts between $-(3.9$-$5.3)\%$, regardless of charge
or flavor, so temperature alone does not distinguish a particle from
its antiparticle. At finite baryon density, a clear
particle-antiparticle asymmetry emerges: at $T = 0$ and $n = 5n_0$,
the $\bar{B}^{*0}$ mass decreases by $12.9\%$, whereas the $B^{*0}$
mass shifts by only $6.1\%$, a gap of nearly seven percentage points
driven entirely by the vector self-energy. This provides a theoretical
basis for the future heavy-ion collision program at RHIC, LHC, FAIR,
and NICA.
\end{abstract}
\maketitle
\section{Introduction}\label{sec:intro}

One of the main objectives of modern high-energy physics is the exploration of the properties of strongly interacting matter in conditions of extreme temperatures and densities.
Such studies are needed to understand the Universe in its early stages, the composition of neutron-star cores, and the matter produced in experiments with relativistic heavy ions~\cite{Shuryak:1978ij,Busza:2018rrf,Annala:2019puf,Baym:2017whm}.
The study of the phase diagram of quantum chromodynamics (QCD) at extreme conditions is also necessary for understanding the strong interaction.

Based on QCD, hadron matter exhibits a phase transition to a deconfined quark-gluon 
plasma (QGP) state at high enough temperature and/or baryon density. The nature of such 
a transition is related to partial restoration of chiral symmetry and 
weakened color confinement~\cite{Braun-Munzinger:2015hba}. Numerical studies of lattice 
QCD at zero baryon density reveal that the crossover transition happens around the 
critical temperature $T_c \sim 150$-$160~\mathrm{MeV}$~\cite{Bazavov:2019lgz}. However, 
the structure of the phase diagram of QCD at nonzero baryon density is yet to be explored.
In particular, the position and nature of the phase transition at finite density are a source of many questions that will be answered by future experiments at FAIR (CBM)~\cite{CBM:2016kpk} 
and NICA (MPD)~\cite{MPD:2022qhn}. The difficulty is finding reliable observables in the hadronic 
sector that will depend on the QCD vacuum even during the collision.

Among the possible hadronic probes of hot and dense QCD matter, vector heavy-light 
mesons such as  $D^*$, $D_s^*$, $B^*$, and $B_s^*$ occupy a particularly 
important position. Since these states contain both heavy and light quarks, they 
simultaneously reflect medium modifications associated with the heavy-quark sector 
and the partial restoration of chiral symmetry in the light-quark sector is expected to carry valuable information about the in-medium behavior of 
hadrons. In particular, the suppression 
of these quantities with increasing temperature and baryon density is commonly 
interpreted as a signal of substantial modifications in the QCD vacuum and the 
gradual restoration of chiral symmetry~\cite{Cobos-Martinez:2025iqg}.

Pioneering work by Hatsuda et al.~\cite{Hatsuda:1996xt} within QCD sum rules (QCDSR) demonstrated that light vector mesons ($\rho$, $\omega$, $\phi$) exhibit mass reductions at finite baryon density, with an estimated shift for $\rho$ and $\omega$ mesons of $m^*/m \simeq 1 - (0.16 \pm 0.06) (\rho/\rho_0)$ at normal nuclear matter density. Within the framework of thermal QCDSR (TQCDSR), it has been observed that while pseudoscalar meson masses exhibit a slight increase of approximately $(10-20)\%$ with temperature, vector meson masses demonstrate a decrease of about $(20-30)\%$ near the critical temperature~\cite{Dominguez:2007}.

In the strange vector sector, Ilner \textit{et al.}~\cite{Ilner:2013ksa}
studied the in-medium properties of the $K^*$ and $\bar{K}^*$ mesons
at finite temperature and baryon chemical potential $\mu_B$ using
chirally motivated meson self-energy models. They found significant
spectral broadening, while the mass shifts remained below $5\%$.
Within the TQCDSR framework, Azizi \textit{et al.}~\cite{Bozkir:2022lyk}
investigated the kaon at finite temperature and baryon chemical
potential. They showed that the kaon mass first increases and then
decreases with increasing $\mu_B$, eventually vanishing near
$\mu_B \simeq (1.03$-$1.15)\,\mathrm{GeV}$. The decay constant also
shows a non-monotonic dependence on $\mu_B$.

The kaon sector provides one of the clearest examples of
particle-antiparticle asymmetry in dense nuclear matter:
the $K^+$ ($u\bar{s}$) meson experiences a repulsive
interaction, whereas the $K^-$ ($\bar{u}s$) meson feels an
attractive potential. This difference originates from the
Weinberg-Tomozawa (WT) term, originally derived for $\pi N$
scattering as the leading-order vector interaction between
a pseudoscalar meson and the baryon-number
current~\cite{Weinberg:1966kf,Tomozawa:1966jm}, and later
extended via SU(3) to the kaon-nucleon channel within a
QCD sum-rule framework~\cite{Cohen:1994wm}, where the opposite-sign vector
potentials felt by $K^+$ and $K^-$ emerge directly from the
quark-density operator entering the
correlation function~\cite{Kondo:1994uw}. As a direct
consequence, this splitting grows monotonically with
increasing baryon density,
reaching $|\Delta m| \sim 0.35~\mathrm{GeV}$ near
$\rho \simeq 3.2\,\rho_{\mathrm{sat}}$ at zero temperature,
and is only partially suppressed by thermal
fluctuations~\cite{Azizi:2026zxh}.

It is therefore natural to ask whether a similar asymmetry appears
in the $B^*$ sector. Experimentally, the $B_s^*(5415)$ and $B^*(5325)$ mesons are well established through high-energy $e^+e^-$ collisions. The charged and neutral $B^*$ states were first observed by the CLEO Collaboration~\cite{CLEO-II:1991pfp}. Later, operating at the $\Upsilon(5S)$ resonance, the same collaboration~\cite{CLEO:2005utd} identified the dominant process $e^+e^- \to B_s^* B_s^*$ and measured its production cross-section as $[0.11^{+0.04}_{-0.03}\,\text{(stat.)} \pm 0.02\,\text{(syst.)}]$~\text{nb}. Unlike the kaon, the mass of the $B^*$ meson is
mainly determined by the heavy $b$ quark, while the light-quark and
gluon condensates are responsible for its in-medium modifications.
An important question is how these two effects compete and whether
the decay constant is more sensitive to the medium than the mass.

Many studies have also explored the in-medium behavior of heavy
vector mesons. Using QCDSR together with the chiral SU(3) model,
Kumar~\cite{Kumar:2014} found that the masses of the $D^*$ and
$B^*$ mesons decrease with increasing temperature and baryon density
because of the reduction of quark and gluon condensates in the
medium. In another work, Kumar and
Mishra~\cite{Kumar:2009xc} studied the $D$ and $\bar{D}$ mesons in
isospin-asymmetric nuclear matter at finite temperature within an
effective chiral SU(4) model. At $\rho_B=4\rho_0$, they found that
the $D^+$ mass decreases by about $20\%$, while the reductions for
the $\bar{D}$ mesons ($D^-$ and $\bar{D}^0$) are around $10\%$.
Their results also show that baryon density has a much stronger
effect than temperature on the in-medium properties of heavy
mesons.

The thermal behavior of heavy quarkonium states has also been studied extensively 
within TQCDSR: Veliev \textit{et al.}~\cite{Veliev:2011kq} calculated the masses 
and decay constants of $J/\psi$ and $\Upsilon$ mesons at finite temperature, 
finding that the decay constants reach approximately $45\%$ of their vacuum values 
at $T_c$, with mass reductions of $12\%$ and $2.5\%$, respectively. 
For light tensor mesons, Azizi \textit{et al.}~\cite{Azizi:2014maa} 
reported even more drastic modifications, with decay constants decreasing by 
$(70$-$85)\%$ and masses by $(60$-$72)\%$ at deconfinement temperature. 
More recently, the properties of $D^*$ mesons in isospin-asymmetric nuclear matter 
were studied as functions of temperature and baryon 
density~\cite{Cobos-Martinez:2025iqg}. It was found that baryon density effects 
dominate over thermal and isospin-asymmetry effects in determining the in-medium 
behavior of the masses and decay constants.

Motivated by these developments, in the present work we investigate the temperature- and density-dependent properties of the $B_s^*$ and $B^*$ vector mesons within the QCD sum rules framework. While the in-medium behavior of open-charm and light vector mesons has been extensively studied, existing sum rule analyses of the beauty sector remain scarce and typically treat temperature and density separately, and rarely confront particles and antiparticles within the same multiplet. Beyond extending previous in-medium studies to the beauty-vector sector under a unified, simultaneous temperature- and density-dependent OPE, our main objective is to provide a systematic comparison of the strange and non-strange, charged and neutral, particle and antiparticle states. This allows us to disentangle the roles of strangeness, light-quark flavor, baryon density, temperature, and charge conjugation in determining the in-medium response of heavy-light vector mesons. In particular, by comparing particle-antiparticle and isospin splittings, we identify the vector self-energy as the dominant source of the charge-conjugation asymmetry in dense non-strange $B^*$ matter and establish its hierarchy relative to the isospin splitting. The comparison with the $B_s^*$ sector further demonstrates that the presence of a strange light quark substantially suppresses this asymmetry. We also find that the decay constants are considerably more sensitive to the medium than the masses. These features provide a more detailed physical picture of the mechanisms governing the response of heavy-light beauty vector mesons to hot and dense QCD matter, and constitute the new physical insight of the present analysis beyond earlier QCD sum-rule studies of heavy-light mesons in medium.

This paper is organized as follows. In Sec.~\ref{sec:framework}, we present the theoretical framework, including the in-medium correlation function and the derivation of the sum rules. Sec.~\ref{sec:numerics} contains the numerical analysis, where we discuss the convergence of the OPE, the determination of the auxiliary parameters, and the temperature- and density-dependent behavior of the physical observables. Finally, our conclusions are summarized in Sec.~\ref{sec:summary}.

\FloatBarrier
\section{Theoretical Framework}\label{sec:framework}

One of the most important analytical techniques for the 
extraction of hadronic observables from the QCD vacuum 
lies within the QCDSR technique proposed by Shifman, 
Vainshtein, and Zakharov~\cite{Shifman:1978bx,
Shifman:1978by}. This technique relies on the OPE of current correlators in the deep Euclidean region and on quark–hadron duality as a matching condition between the QCD and hadronic representations. The extension of QCD sum rules to finite temperature was first developed by Bochkarev and Shaposhnikov~\cite{Bochkarev:1985ex}. Further developments and applications at finite temperature and density can be found in Ref.~\cite{Mallik:1997kj}. The in-medium 
modifications of hadron properties at finite 
temperature~\cite{Mallik:1997kj,
Ayala:2016vnt,Veliev:2011kq,
Sungu:2020zvk,Aydin:2025lbl,Azizi:2019kzj,
Sungu:2020azn}, at finite baryon 
density~\cite{Cohen:1991nk,Azizi:2020lnh,Azizi:2018gsp,
Azizi:2016dhy,Azizi:2014yla,Azizi:2022jzl,
Er:2024eph}, and under the simultaneous effects of 
both temperature and baryon density~\cite{Cohen:1994wm,Adami:1993tp,Azizi:2026zxh,
Tang:2011za,Kim:2017nyg,Glozman:2022zpy,Kumar:2009xc} 
have been extensively investigated in the literature 
within various theoretical frameworks, providing 
valuable insights into the QCD phase structure and 
the partial restoration of chiral symmetry in hot and 
dense nuclear matter.
%
\subsection{Two-Point Correlator in Nuclear Matter 
at Finite \texorpdfstring{$T$}{T}}
The analysis is based on the two-point 
correlation function evaluated in a medium of nuclear 
matter at temperature $T$ and baryon number density $n$,
\begin{equation}\label{eq:CF}
\Pi_{\mu\nu}(p,T,n)
= i\!\int\!d^4x\,e^{ip\cdot x}
\langle\mathcal{T}\{J_\mu(x)J_\nu^\dagger(0)\}
\rangle_{T,n},
\end{equation}
where the interpolating vector current for a meson with 
quark content $\bar{\psi}_1\psi_2$ is
\begin{equation}\label{current}
J_\mu(x) = \bar{\psi}_1(x)\,\gamma_\mu\, \psi_2(x).
\end{equation}

The quark content of the $B_s^*(5415)$
and  $B^*(5325)$  mesons studied in this work is 
summarized in Table~\ref{tab:meson_properties}.
\begin{table}[H]
\centering
\renewcommand{\arraystretch}{1.5}
\caption{\justifying Quark content of the   $B_s^*(5415)$
and $B^*(5325)$ mesons and their 
antiparticles~\cite{PDG:2024}.}
\label{tab:meson_properties}
\setlength{\tabcolsep}{12pt}
\begin{tabular}{lc}
\hline\hline
Meson & Quark Content \\
\hline
$B^{*+}$            & $\bar{b}u$ \\
$B^{*0}$            & $\bar{b}d$ \\
$B_s^{*0}$          & $\bar{b}s$ \\
\hline
$B^{*-}$            & $b\bar{u}$ \\
$\bar{B}^{*0}$      & $b\bar{d}$ \\
$\bar{B}_s^{*0}$    & $b\bar{s}$ \\
\hline\hline
\end{tabular}
\end{table}
The expectation value $\langle\cdots\rangle_{T,n}$ is taken over the grand canonical ensemble at finite temperature and baryon density. The presence of the medium explicitly breaks Lorentz invariance, leaving only spatial rotational symmetry $SO(3)$ in the rest frame defined by the four-velocity $u^\mu=(1,\mathbf{0})$. The $B_s^*$  and $B^*$ meson properties are extracted 
through a dual evaluation of $\Pi_{\mu\nu}(p,T,n)$. 
On the hadronic side, we invoke the physical spectral 
representation, while on the QCD side, we employ the 
OPE in the short-distance limit. The two representations 
are matched via a dispersion relation, after which a 
Borel transformation and continuum subtraction are 
applied. This procedure ensures that the sum rules are 
dominated by the ground-state contribution, effectively 
suppressing higher resonances and continuum states.
\subsection{Hadronic representation}

In the physical representation, the correlation function is constructed using hadronic degrees of freedom. After performing the four-dimensional spacetime integration, we obtain
\begin{equation}\label{eq:PR1}
\Pi_{\mu\nu}^{\rm Had}(p,T,n)
= -\frac{\langle J_\mu | B(p) \rangle \langle B(p) | J{\dagger}_{\mu} \rangle}{p^{*2}-\tilde{m}^{2}(T,n)}
  + \cdots,
\end{equation}
where $p^{*}$ denotes the in-medium momentum and $\tilde{m}(T,n)$ represents the effective mass of the $B$ state, modified by the effects of the hot and dense medium. The ellipsis (...) accounts for the contributions from higher resonances and the continuum spectrum.

The modified decay constant, or decay constant, $\tilde{f}(T,n)$ is defined via the following matrix element in terms of the polarization vector $\epsilon_\mu$ of the $B$ state:
\begin{equation}\label{eq:PR2}
\langle J_\mu | B(p)\rangle_{T,n}
= \tilde{f}^(T,n)\,\tilde{m}(T,n)\,\epsilon_\mu.
\end{equation}
In what follows, we shall use the notation $\tilde{m}$ and $\tilde{f}$ to denote $\tilde{m}(T,n)$ and $\tilde{f}(T,n)$, respectively. 

Inserting Eq.~\eqref{eq:PR2} into Eq.~\eqref{eq:PR1} and summing over the polarization vectors, the hadronic side of the 
correlation function takes the form
\begin{equation}\label{eq:PR3}
\Pi^{\rm Had}_{\mu\nu}(p,T,n)
= \frac{\tilde{f}^2\,\tilde{m}^2}{p^{*2}-\tilde{m}^2}
  \left( -g_{\mu\nu} + \frac{p^*_{\mu} p^*_{\nu}}{\tilde{m}^2} \right)
  + \cdots.
\end{equation}
The in-medium four-momentum $p^*_\mu$ encodes the interaction of the 
meson with the nuclear matter through the vector self-energy 
$\Sigma_{\mu,v}$~\cite{Cohen:1994wm},
\begin{equation}\label{eq:PR4}
p^*_{\mu} = p_{\mu} - \Sigma_{\mu,v}, \qquad
\Sigma_{\mu,v} = \Sigma_{v}\, u_{\mu} + \Sigma'_{v}\, p_{\mu},
\end{equation}
where $\Sigma_v$ denotes the vector self-energy of the $B$ meson and 
$u_\mu$ is the four-velocity of the nuclear medium. Within the 
mean-field approximation, both the scalar and vector self-energies are 
taken to be real and momentum-independent, and $\Sigma'_v$ vanishes 
identically. Working in the rest frame of the medium, 
$u_\mu = (1,\boldsymbol{0})$, and substituting Eq.~\eqref{eq:PR4} 
into Eq.~\eqref{eq:PR3}, the hadronic representation becomes
\begin{align}\label{eq:PR5}
\Pi^{\rm Had}_{\mu\nu}(p,T,n)
=& -\frac{\tilde{f}^2}{p^2 - \mu^2}
\Big[-g_{\mu\nu}\tilde{m}^2 + p_{\mu} p_{\nu}          \notag\\
  &- \Sigma_{v}(p_{\mu} u_{\nu} + p_{\nu} u_{\mu})
   + \Sigma_{v}^2\, u_{\mu} u_{\nu}
\Big] + \cdots,
\end{align}
here the shifted pole mass is defined through 
$\mu^2 \equiv \tilde{m}^2 - \Sigma_v^2 + 2p_0\Sigma_v$, with 
$p_0 = p\cdot u$ the quasi-particle energy. 

The suppression of higher-state and continuum contributions is 
achieved by applying the Borel transformation operator, defined 
through the following limiting procedure:
\begin{equation}\label{eq:PR6}
\hat{\mathcal{B}} = \lim_{\substack{Q^2, m \rightarrow \infty \\ Q^2/m = M^2}} \frac{(-p^2)^m}{(m-1)!} \left( \frac{d}{dp^2} \right)^{m-1},
\end{equation}
where $M^2$ denotes the Borel mass parameter. The action of $\hat{\mathcal{B}}$ on some typical functions relevant to the present calculation is summarized as follows:
\begin{eqnarray}\label{eq:PR7}
\hat{\mathcal{B}} e^{-\alpha p^2} &= &\delta( 1/M^2-\alpha), \nonumber \\
\quad \hat{\mathcal{B}} \left( \frac{1}{(m^2 - p^2)^k} \right) &=& \frac{1}{(k-1)! (M^2)^{k-1}} e^{-m^2/M^2}.
\end{eqnarray}

Applying $\hat{\mathcal{B}}$ to Eq.~\eqref{eq:PR5} with respect to $Q^2 \equiv -p^2$, and utilizing the identities in Eq.~\eqref{eq:PR7}, we obtain the Borel-transformed hadronic representation:
\begin{eqnarray}
 \mathbf{\Pi}^{\rm Had}_{\mu\nu}(M^2,T,n)
&= &\tilde{f}^2\, e^{-\mu^2/M^2}
\Big[-g_{\mu\nu}\tilde{m}^2 + p_{\mu} p_{\nu} \nonumber\\
  &-& \Sigma_{v}(p_{\mu} u_{\nu} + p_{\nu} u_{\mu})
   + \Sigma_{v}^2\, u_{\mu} u_{\nu}
\Big]\nonumber\\&+& \cdots.
\end{eqnarray}\label{eq:PR8}
\vspace{-14pt}
\subsection{QCD representation}

The QCD side of the correlation function is obtained by substituting 
the interpolating current $J_\mu(x)$, defined in Eq.~\eqref{current}, 
into Eq.~\eqref{eq:CF} and contracting all quark-field pairs via 
Wick's theorem. The resulting expression is expressed in terms of the in-medium light- and heavy-quark propagators as
\begin{equation}\label{eq:QCD1}
\Pi_{\mu\nu}^{\rm QCD}(p,T,n)
= i \int d^4x\, e^{ip\cdot x}\,
  \mathrm{Tr}\!\left[
    S^{ab}_{\psi_1}(x)\,\gamma_{\nu}\,
    S^{ba}_{\psi_2}(-x)\,\gamma_{\mu}
  \right],
\end{equation}
where $S^{ab}_{\psi_{1,2}}(x)$ denote the full quark propagators in a hot and dense medium, with $a$ and $b$ representing the color indices. The specific quark flavor assignments, $\psi_1$ and $\psi_2$, corresponding to each meson state are summarized in Table~\ref{tab:meson_properties}.

In  the fixed-point 
gauge, the coordinate-space representation of the full 
in-medium light-quark ($q = u, d, s$) propagator 
$S^{ab}_{q}(x)$ at finite temperature and baryon density 
reads
\begin{align}\label{eq:lightprop}
S_q^{ab}(x)
  &= \left(1 - n_F\right)
     \left[
       \frac{i\slashed{x}}{2\pi^2 x^4}
       - \frac{m_q}{4\pi^2 x^2}
     \right]\delta^{ab}
     + \langle \chi_q^a(x)\,\bar{\chi}_q^b(0) \rangle
     \notag\\
  &\quad
     - \frac{ig_s}{32\pi^2}\,
       \frac{\slashed{x}\sigma_{\mu\nu}
             + \sigma_{\mu\nu}\slashed{x}}{x^2}\,
       G^{\mu\nu}_A(0)\,t_A^{ab}
     \notag\\
  &\quad
     + \frac{i}{3}\left[
         -\frac{\slashed{x}}{12}
         + \frac{1}{3}(u\cdot x)\slashed{u}
       \right]
       \langle u^{\mu}\Theta^f_{\mu\nu}u^{\nu}\rangle\,
       \delta^{ab}
     + \cdots,
\end{align}
and the full in-medium heavy-quark ($Q = b$) propagator 
$S_Q^{ab}(x)$ is
\begin{align}\label{eq:heavyprop}
S_Q^{ab}(x)
  &= \frac{i}{(2\pi)^4}\int d^4k\, e^{-ik\cdot x}
     \Bigg\{
       \left(1 - n_F\right)
       \frac{\delta_{ab}}{\slashed{k} - m_Q}
     \notag\\
  &\quad
       - \frac{g_s\, G_{\mu\nu}^{ab}(0)}{4}\,
         \frac{
           \sigma^{\mu\nu}(\slashed{k}+m_Q)
           +(\slashed{k}+m_Q)\sigma^{\mu\nu}
         }{(k^2-m_Q^2)^2}
     \notag\\
  &\quad
       + \frac{\pi^2}{3}
         \left\langle\frac{\alpha_s}{\pi}G^2\right\rangle
         \delta_{ab}\, m_Q\,
         \frac{k^2 + m_Q\slashed{k}}{(k^2-m_Q^2)^4}
       + \cdots
     \Bigg\}.
\end{align}
In both propagators, the medium effects enter through the 
Fermi-Dirac distribution function
\begin{equation}\label{eq:FD}
n_F = \frac{1}{e^{\,|p_0|/T}+1},
\end{equation}
where $p_0$ is the energy component of the four-momentum 
$p^\mu = (p_0, \mathbf{p})$. The factor $(1-n_F)$ accounts 
for Pauli blocking of occupied fermionic states, and both 
propagators reduce to their standard vacuum forms in the 
limit $T \to 0$, $n \to 0$. The gluon and quark condensates 
appearing in the propagators carry an additional medium 
dependence through both $T$ and $n$~\cite{Kumar:2014}. 
For a detailed derivation of the nonperturbative contributions to the propagator, we refer the reader to Ref.~\cite{Bozkir:2022lyk} for hot medium effects and Ref.~\cite{Er:2022cxx} for dense matter conditions.

In Eqs.~\eqref{eq:lightprop} and~\eqref{eq:heavyprop}, 
$\chi_q^a$ and $\bar{\chi}_q^b$ are the Grassmann-valued 
background quark fields whose vacuum expectation values yield 
the quark condensates, and $G^{\mu\nu}_A$ is the classical 
background gluon field-strength tensor. The quantity 
$\langle u^{\mu}\Theta^f_{\mu\nu}u^{\nu}\rangle$ appearing 
in Eq.~\eqref{eq:lightprop} denotes the thermal expectation 
value of the fermionic part of the energy-momentum tensor, 
projected along the medium four-velocity $u^\mu$.
\vspace{-21pt}
\subsection{Condensates at Finite \texorpdfstring{$T$}{T}
and \texorpdfstring{$n$}{n}}
A fundamental input for any QCDSR calculation at finite 
temperature and density is the parametrization of the in-medium condensates. 
In this study, the temperature and nuclear density dependence of the quark 
and gluon condensates are based on the framework of Kumar~\textit{et al.}~\cite{Kumar:2014}, 
who computed these modifications numerically within the chiral $SU(3)$ 
model~\cite{Papazoglou:1998vr}. For the effective continuum threshold $s_0(T,n)$, 
we employ a physically motivated prescription implemented via the Hilbert-moment 
formula of Dominguez, Loewe, and Rojas~\cite{Dominguez:2007} for heavy-light mesons.\\

\noindent\textbf{\textit{Modified quark condensate:}}
To describe the medium modification of the light-quark condensate, we have 
constructed a unified analytic parametrization based on the numerical results of 
Ref.~\cite{Kumar:2014}. Since the original work presents these modifications only 
graphically, we performed an analytical fit to their data to obtain the 
following closed-form expressions as functions of $T$ and $n$:
\vspace{8pt}
\begin{equation}\label{eq:qcKumar}
\frac{\langle\bar{q}q\rangle(T,n)}{\langle\bar{q}q\rangle_0}
= \exp\!\left[
  -\frac{A\,\xi}{1+B\,\xi^\kappa}
  + D\,\xi^\phi\!\left(\frac{T}{T_c}\right)^{\!\alpha}
  - F\!\left(\frac{T}{T_c}\right)^{\!\beta}
  \right],
\end{equation}
where $\xi = n/n_0$ is the baryon number density normalized to the nuclear 
saturation density $n_0 \simeq 0.16~\mathrm{fm}^{-3}$, and the fit 
parameters are listed in Table~\ref{tab:kumar_quark}. 

\begin{table}[H]
\centering
\caption{\justifying Parameters  in Eq.~\eqref{eq:qcKumar}.}
\label{tab:kumar_quark}
\begin{tabular}{cc}
\hline\hline
Symbol                       & Value \\
\hline
$\langle\bar{q}q\rangle_0$   & $-1.4014\times10^{-2}~\mathrm{GeV}^3$ \\
$A$                          & $0.5316$  \\
$B$                          & $0.1370$  \\
$\kappa$                     & $1.2262$  \\
$D$                          & $0.1345$  \\
$\phi$                       & $0.3374$  \\
$\alpha$                     & $1.2516$  \\
$F$                          & $0.0554$  \\
$\beta$                      & $20.0$    \\
\hline\hline
\end{tabular}
\end{table}

In the vacuum limit:
\begin{equation}\label{eq:limits}
\xi \to 0, \quad T \to 0 \quad \Rightarrow \quad
\frac{\langle\bar{q}q\rangle(T,n)}{\langle\bar{q}q\rangle_0} \to 1,
\end{equation}
so that Eq.~\eqref{eq:qcKumar} correctly reduces to $\langle\bar{q}q\rangle_0$.

In the isospin-symmetric nuclear matter, the number-density condensates 
satisfy
\begin{equation}
\langle q^\dagger q\rangle_{T,n}=\frac{3}{2}n, \qquad
\langle s^\dagger s\rangle_{T,n}=0,
\end{equation}
as established in Refs.~\cite{Cohen:1991nk,Er:2022cxx}.

\FloatBarrier
\medskip
\noindent\textbf{\textit{Modified gluon condensate:}}
The finite-temperature and finite-density modification of the gluon 
condensate is likewise parametrized using an analytical fit constructed in 
this work, based on the numerical results of Kumar~\textit{et al.}~\cite{Kumar:2014}. 
In their work, the gluon condensate was evaluated within the chiral $SU(3)$ 
model~\cite{Papazoglou:1998vr,Kumar:2009xc} incorporating constraints from 
the lattice QCD data of Ref.~\cite{Kwon:2010fw}. To obtain a practical 
closed-form expression, we have fitted their numerical data, yielding:
\begin{align}\label{eq:G2Kumar}
\left\langle\frac{\alpha_s}{\pi}G^2\right\rangle\!(T,n)
  &= A_0 - \frac{\xi}{T_c(1+\xi)^2}
     \Big[A_5\,T\,(1+\xi) \nonumber\\
  &+ T_c\big((2+\xi)\,A_3
                  +(1+\xi)\,A_4\big)\Big],
\end{align}
where the dimensionless density parameter $\xi$ is defined as above, 
and the fitting parameters are listed in Table~\ref{tab:kumar_gluon}. 
The vacuum gluon condensate $A_0 \equiv \langle\frac{\alpha_s}{\pi}G^2\rangle_0$ 
is recovered in the limit
\begin{equation}
\xi \to 0, \quad T \to 0 \quad \Rightarrow \quad
\left\langle\frac{\alpha_s}{\pi}G^2\right\rangle \to A_0.
\end{equation}
\vspace{-20pt}
\begin{table}[H]
\centering
\caption{\justifying Parameters of the gluon condensate fit in Eq.~\eqref{eq:G2Kumar}.}
\label{tab:kumar_gluon}
\begin{tabular}{cc}
\hline\hline
Symbol & Value ($\mathrm{GeV}^4$) \\
\hline
$A_0$  & $\phantom{-}0.0195\phantom{000}$ \\
$A_3$  & $-4.131\times10^{-3}$            \\
$A_4$  & $\phantom{-}7.696\times10^{-3}$  \\
$A_5$  & $-8.530\times10^{-4}$            \\
\hline\hline
\end{tabular}
\end{table}

\noindent\textbf{\textit{Fermionic energy-momentum tensor:}}
The in-medium light-quark propagator in Eq.~\eqref{eq:lightprop}  contains the thermal matrix element  $\langle u^\mu\Theta^f_{\mu\nu}u^\nu\rangle$, which encodes the 
contribution of quark degrees of freedom to the energy density of the medium. Assuming equal partitioning between the quark and gluon  sectors~\cite{Shuryak:1993kg,Mallik:1997kj};

\begin{equation*}
\langle\Theta_{00}^g\rangle = \langle\Theta_{00}^f\rangle 
= \frac{1}{2}\langle\Theta_{00}\rangle.
\end{equation*}
With this approximation, we use the fit function from Ref.~\cite{Azizi:2015ona} for the fermionic and gluonic contributions to the energy-momentum tensor, based on the lattice QCD results of Ref.~\cite{Cheng:2007jq}:

\begin{align}
\langle\Theta_{00}^g\rangle = \langle\Theta_{00}^f\rangle 
&= T^4\, \exp\!\left[113.867\left(\frac{T}{\mathrm{GeV}}\right)^{\!2}\right.
\nonumber\\
& \left.- 12.190\left(\frac{T}{\mathrm{GeV}}\right)\right]
- 10.141\left(\frac{1}{\mathrm{GeV}}\right)T^5.
\end{align}
\medskip
\noindent\textbf{\textit{Modified mixed quark-gluon condensate:}}
In nuclear matter, the mixed condensate receives both a thermal
contribution and a density-driven correction~\cite{
Er:2022cxx},
\begin{align}
\langle\bar{q}g_s\sigma G q\rangle_{T,n} &= m_0^2\,\langle\bar{q}q\rangle(T,n)
  + 3\, n\;\mathrm{GeV}^2, \nonumber \\
\langle\bar{s}g_s\sigma G s\rangle_{T,n} &= m_0^2\,\langle\bar{s}s\rangle(T,n)
  + 3\,y\, n\;\mathrm{GeV}^2,
\end{align}
with $m_0^2=(0.8\pm0.1)\;\mathrm{GeV}^2$~\cite{Reinders:1984sr}
and $y=0.05$~\cite{Er:2022cxx}.

The higher-order nuclear corrections to the vector-mixed
and derivative condensates read~\cite{Er:2022cxx}
\begin{align}
\langle q^\dagger g_s\sigma G q\rangle(T,n)
  &= -0.33\; n\;\mathrm{GeV}^2, \nonumber \\
\langle s^\dagger g_s\sigma G s\rangle(T,n)
  &= -0.33\; y\, n\;\mathrm{GeV}^2, \nonumber \\
\langle\bar{q}\,iD_0 iD_0 q\rangle(T,n)
  &= 0.3\; n\;\mathrm{GeV}^2
     -\tfrac{1}{8}\langle\bar{q}g_s\sigma G q\rangle(T,n), \nonumber \\
\langle\bar{q}\,iD_0 q\rangle(T,n)
  &= \tfrac{3}{2}\,m_q\, n, \nonumber \\
\langle\bar{s}\,iD_0 s\rangle(T,n)
  &= 0, \nonumber \\
\langle q^\dagger\,iD_0 iD_0 q\rangle(T,n)
  &= 0.031\; n\;\mathrm{GeV}^2
     -\tfrac{1}{12}\langle q^\dagger g_s\sigma G q\rangle(T,n), \nonumber \\
\langle s^\dagger\,iD_0 iD_0 s\rangle(T,n)
  &= 0.031\; y\, n\;\mathrm{GeV}^2
     -\tfrac{1}{12}\langle s^\dagger g_s\sigma G s\rangle(T,n), \nonumber \\
\langle q^\dagger\,iD_0 q\rangle(T,n)
  &= 0.18\; n, \nonumber \\
\langle s^\dagger\,iD_0 s\rangle(T,n)
  &= m_s\,\langle\bar{s}s\rangle(T,n) + 0.02\; n.
\end{align}

\FloatBarrier
\subsection{Effective Continuum Threshold}
\label{subsec:s0}

The effective continuum threshold $s_0(T,n)$ encodes the
separation between the ground-state pole and the hadronic continuum.
In the vacuum, it is fixed phenomenologically near the squared mass of
the first excited state of each meson. At finite $T$ and $n$ it
must decrease toward the perturbative QCD value, reflecting the
gradual dissolution of hadronic bound states as the density and
temperature increase.

The medium modification of $s_0$ depends on whether light or
heavy-quark states are considered.
For heavy-light systems, the heavy quark mass $m_Q$ sets a
perturbative QCD floor for the threshold. Following
Dominguez, Loewe, and Rojas~\cite{Dominguez:2007}, we write:
\begin{align}\label{eq:s0heavy}
\frac{s_0(T,n)}{s_0}
&= \frac{\langle\bar{q}q\rangle(T,n)}
        {\langle\bar{q}q\rangle_0}
   \left[1-\frac{m_Q^2}{s_0}\right]+ \frac{m_Q^2}{s_0}.
\end{align}
In the chiral-restoration limit, where the in-medium 
quark condensate vanishes, 
$|\langle\bar{q}q\rangle(T,n)|\to 0$ as 
$T\to T_c$ and $n\to n_0$, the ratio 
$\langle\bar{q}q\rangle(T,n)/\langle\bar{q}q\rangle_0 
\to 0$, and consequently the first term in 
Eq.~\eqref{eq:s0heavy} vanishes. In this limit, 
$s_0(T,n)\to m_Q^2$, so that the continuum threshold 
decreases toward the heavy-quark mass squared, which 
sets the kinematic lower bound for meson production.

\FloatBarrier
\subsection{Sum Rules}
\label{subsec:sumrules}
In the QCD representation of the correlation function, the coordinate-space expression is derived by employing the aforementioned medium-dependent operators. To transform the calculations into momentum space, we perform a Fourier transformation as follows:
\begin{eqnarray}
\label{ }
\frac{1}{(x^2)^m}&=&\int \frac{d^D t }{(2\pi)^D}e^{-i t \cdot x}i(-1)^{m+1}2^{D-2m}\pi^{D/2} \nonumber \\
&\times& \frac{\Gamma[D/2-m]}{\Gamma[m]}\Big(-\frac{1}{t^2}\Big)^{D/2-m}.
\end{eqnarray}
By substituting $x_{\nu}$ with $-i\frac{\partial}{\partial p_{\nu}}$, the remaining four-dimensional integrals are evaluated using the following identity:
\begin{equation}
\label{ }
\int d^4 \ell\frac{(\ell^2)^m}{(\ell^2+\textsf{L})^n}=\frac{i\pi^2 (-1)^{m-n} \Gamma[m+2]\Gamma[n-m-2]}{\Gamma[2]\Gamma[n] (-\textsf{L})^{n-m-2}}.
\end{equation} 
Using the following relation, we extract the imaginary parts corresponding to the various structures:
\begin{equation}
\label{ }
\Gamma\Big[\frac{D}{2}-n\Big]\Big(-\frac{1}{\textsf{L}}\Big)^{D/2-n}=\frac{(-1)^{n-1}}{(n-2)!}(-\textsf{L})^{n-2}ln[-\textsf{L}].
\end{equation} 

The results obtained depend on the continuum threshold $s_0$ and the Borel parameter $M^2$. By applying the continuum subtraction procedure to suppress further the contributions from higher excited states and the continuum, we arrive at the following representation in the Borel scheme:
\begin{equation}
\label{ }
\Pi_{i}^{QCD}(s_0,M^2,T,n)=\int_{(m_q+m_Q)^2}^{s_0(T,n)} ds \rho_i^{QCD} (s) e^{\frac{-s}{M^2}}.
\end{equation}

Here,   $\rho_i^{QCD} (s)$, with $i$ denoting different structures, are the new spectral densities, and they are given in terms of the perturbative (pert)  and non-perturbative (non-pert) components,
\begin{equation}
\label{ }
\rho^{QCD} (s)=\rho^{pert} (s)+\rho^{non-pert } (s),
\end{equation}
where $\rho^{non-pert } (s)$ contains quark, gluon  and mixed quark-gluon  condensates. As an example, the explicit expression for the QCD side associated with the $g_{\mu\nu}$ structure is provided in the Appendix.

By equating the coefficients of the chosen Lorentz structures from both the hadronic and QCD representations of the correlation function, we arrive at the following sum rules:
\begin{eqnarray}\label{eq:SumRules}
-\tilde{m}^2 \tilde{f}^2 e^{-\mu^2/M^2}& = & \Pi^{QCD}_{g_{\mu\nu}}, \nonumber  \\
\tilde{f}^2 e^{-\mu^2/M^2}& = & \Pi^{QCD}_{p_{\mu} p_{\nu}}, \nonumber  \\
-\Sigma_{\upsilon} \tilde{f}^2 e^{-\mu^2/M^2}& = & \Pi^{QCD}_{p_{\mu} u_{\nu}}, \nonumber  \\
-\Sigma_{\upsilon} \tilde{f}^2 e^{-\mu^2/M^2}& = & \Pi^{QCD}_{p_{\nu} u_{\mu}}, \nonumber  \\
\Sigma^2_{\upsilon} \tilde{f}^2 e^{-\mu^2/M^2}& = & \Pi^{QCD}_{u_{\mu} u_{\nu}},
\end{eqnarray}
which will be employed in our numerical analysis.

\section{Numerical Analysis}\label{sec:numerics}
This section is devoted to a detailed numerical study of 
how the masses and decay constants of the $B_s^*$ and  $B^*$
 multiplets change when these mesons are placed 
in a hot and dense nuclear matter. Rather than 
treating a single representative state, we follow all 
six charge-flavor configurations explicitly,
$B_s^*\,(\bar{b}s)$, $\bar{B}_s^*\,(b\bar{s})$, 
$B^{*+}\,(\bar{b}u)$, $B^{*-}\,(b\bar{u})$, 
$B^{*0}\,(\bar{b}d)$, and $\bar{B}^{*0}\,(b\bar{d})$,  so that the role of the quark content in shaping 
the medium response can be read off directly from the 
results. 
\begin{table}[H]
\centering
\caption{\justifying QCD input parameters used in the numerical
         analysis.}
\label{tab:params}
\setlength{\tabcolsep}{6pt}
\renewcommand{\arraystretch}{1.3}
\begin{tabular}{lcc}
\hline\hline
Parameter & Value & Unit \\
\hline
\multicolumn{3}{l}{\textit{Quark masses $(\overline{\mathrm{MS}})$ scheme}} \\
\hline
$m_u$ & $2.16^{+0.49}_{-0.26}$ & MeV~\cite{PDG:2024} \\
$m_d$ & $4.67^{+0.48}_{-0.17}$ & MeV~\cite{PDG:2024} \\
$m_s$ & $93.4^{+8.6}_{-3.4}$   & MeV~\cite{PDG:2024} \\
$m_b$ & $4.18^{+0.03}_{-0.02}$ & GeV~\cite{PDG:2024} \\
$m_q$ & $0.5(m_u+m_d)$ & MeV~\cite{PDG:2024}\\
\hline
\multicolumn{3}{l}{\textit{Condensates}} \\
\hline
$\langle\bar{u}u\rangle_0$
      & $-(0.24\pm0.01)^3$  
      & GeV$^3$~\cite{Shifman:1978bx} \\
$\langle\bar{s}s\rangle_0$
      & $0.8\,\langle\bar{u}u\rangle_0$  
      & GeV$^3$~\cite{Narison:2010cg} \\
$\langle\alpha_s G^2\rangle_0$
      & $0.020\pm0.004$  
      & GeV$^4$~\cite{Narison:2010cg} \\
$m_0^2$
      & $0.80\pm0.10$  
      & GeV$^2$~\cite{Reinders:1984sr} \\
\hline
\multicolumn{3}{l}{\textit{Medium scales}} \\
\hline
$T_c$ & $155$ & MeV~\cite{Bazavov:2019lgz} \\
$n_0$ & $0.16$ & fm$^{-3}$~\cite{Cohen:1991nk} \\
\hline\hline
\end{tabular}
\end{table}
\vspace{-25mm}
The comparison between the strange and 
non-strange channels is of particular interest, since 
it gives access to the net effect of the strange-quark 
mass and the associated condensates on the in-medium 
spectral parameters.

The QCD input parameters entering the calculation are 
collected in Table~\ref{tab:params}. All quark masses 
are given in the $\overline{\mathrm{MS}}$ scheme, with 
the renormalization scale chosen as 
$\mu_R = 2\;\mathrm{GeV}$ in the light sector and 
$\mu_R = m_Q$ for the bottom quark. Alongside the QCD inputs, the sum rules derived in Sec.~\ref{subsec:sumrules} 
depend on two auxiliary parameters: the Borel parameter $M^2$ and the effective 
continuum threshold $s_0(T,n)$. These quantities carry no direct physical 
meaning and must therefore be carefully constrained before extracting any 
physical observables. 

The standard strategy~\cite{Shifman:1978bx, Reinders:1984sr} is to identify a 
working window in the $M^2$-$s_0$ plane where the predicted hadronic properties 
are minimally sensitive to variations in both parameters, subject to the 
standard self-consistency criteria. Imposing these requirements in the 
vacuum limit ($T=0$, $n=0$) yields the following working intervals:
\vspace{-25mm}
\begin{equation}
\label{eq:BorelWindow}
\begin{aligned}
  M^2 &\in [12,\;16]\;\mathrm{GeV}^2, \\
  s_0 &\in [33,\;35]\;\mathrm{GeV}^2.
\end{aligned}
\end{equation}
\begin{figure}[H]
 \centering
 \hspace*{-0.1\columnwidth}
  \includegraphics[width=0.52\textwidth]{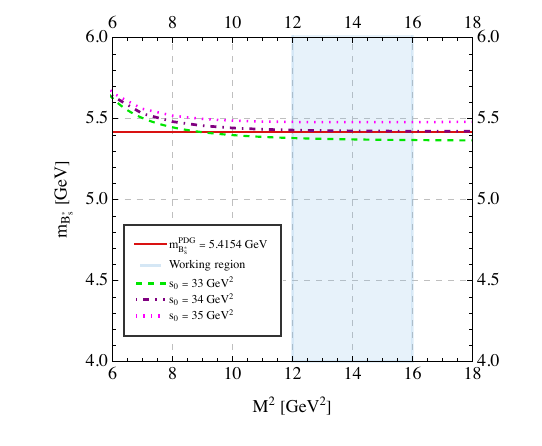}
  \caption{\justifying The vacuum mass of the $B_s^*$ meson, $m_{B_s^*}$, as a 
  function of the Borel parameter $M^2$ for three representative values of 
  the effective continuum threshold $s_0$ (specifically, $s_0 = 33, 34,$ and $35~\mathrm{GeV}^2$). 
  The stability of the plateau curves within the designated interval $M^2 \in [12, 16]~\mathrm{GeV}^2$ 
  illustrates the reliability of the sum rule extraction.}
  \label{fig:Borel}
\end{figure}

\begin{figure}[H]
  \centering
  \hspace*{-0.1\columnwidth}
  \includegraphics[width=0.52\textwidth]{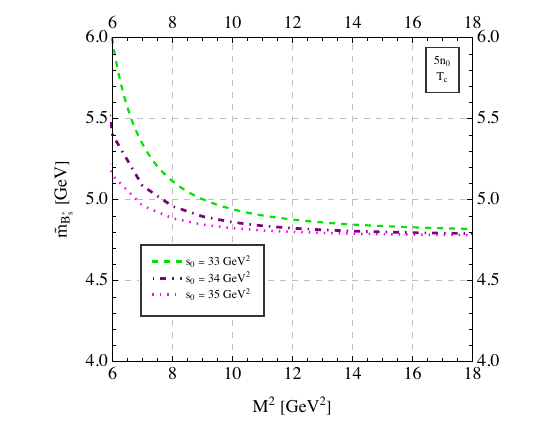}
  \caption{\justifying In-medium mass $\tilde{m}_{B_s^*}$ as a function of the Borel
parameter at $n = 5n_0$ and $T = T_c$, for the same
representative values of the continuum threshold $s_0$ used in
Fig.~\ref{fig:Borel}.}
  \label{fig:BorelHD}
\end{figure}
Since the masses of the vector mesons under investigation are relatively close 
to one another, we illustrate the Borel stability analysis explicitly for the 
$B_s^*$ state in Fig.~\ref{fig:Borel}; the remaining states exhibit qualitatively 
identical stability profiles within the same window. Consequently, we conclude 
that these chosen intervals guarantee a robust and reliable extraction of the 
spectroscopic parameters across the entire multiplet.

\begin{figure}[H]
    \centering
    \hspace*{-0.15\columnwidth}
    \begin{subfigure}[b]{\columnwidth}
        \centering
        \includegraphics[width=1.05\columnwidth]{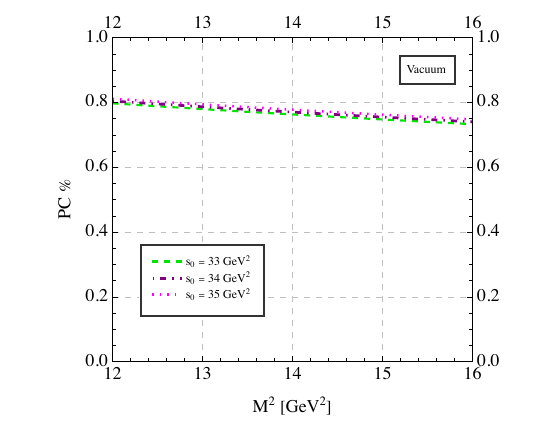}
        \caption{Vacuum}
        \label{fig:PC_vacuum}
    \end{subfigure}
    \vspace{0.15cm}
    \begin{subfigure}[b]{\columnwidth}
        \centering
        \hspace*{-0.15\columnwidth}
        \includegraphics[width=1.05\columnwidth]{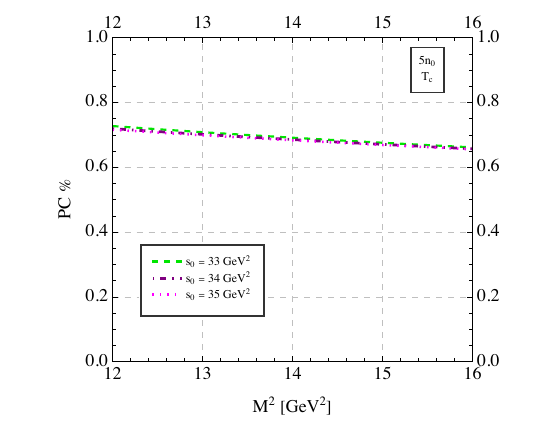}
        \caption{$n = 5n_0, T = T_c$}
        \label{fig:PC_extreme}
    \end{subfigure}
    \caption{\justifying The pole contribution ($\mathrm{PC}\%$) for the $B_s^*$ meson as a function of $M^2$ in vacuum (a) and under extreme medium conditions at $n = 5n_0$ and $T = T_c$ (b) for the specified $s_0$ values.}
    \label{fig:pole_continuation}
\end{figure}

Furthermore, to examine the stability of the sum rules analysis under extreme medium conditions, we present the Borel curves for the $B_s^*$ meson at $n=5n_0$ and $T=T_c$ in Fig.~\ref{fig:BorelHD}. The results show that the extracted mass remains stable over a broad range of the Borel parameter for the considered values of the continuum threshold $s_0$. This behavior indicates that the Borel stability of the sum rules is preserved even under strong thermal and density effects, supporting the reliability of the mass extraction throughout the considered phase diagram.

\begin{figure}[H]
    \centering
    \begin{subfigure}[b]{\columnwidth}
     \hspace*{-0.15\columnwidth}
        \centering
        \includegraphics[
            width=1.05\columnwidth,
            trim=0 0 0 0,
            clip
        ]{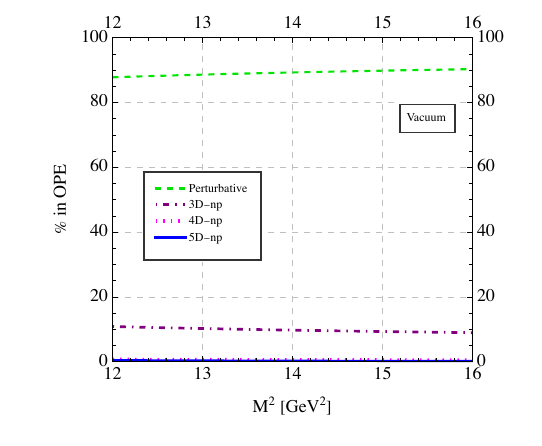}
        \caption{Vacuum}
        \label{fig:OPE_vacuum}
    \end{subfigure}
    \vspace{0.15cm}
    \begin{subfigure}[b]{\columnwidth}
     \hspace*{-0.15\columnwidth}
        \centering
        \includegraphics[
            width=1.05\columnwidth,
            trim=0 0 0 0,
            clip
        ]{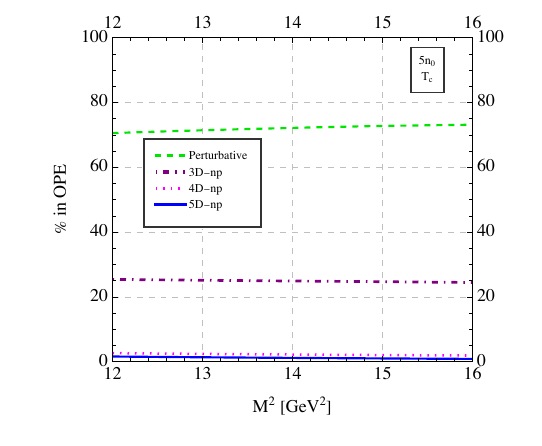}
        \caption{$n = 5n_0, T = T_c$}
        \label{fig:OPE_extreme}
    \end{subfigure}
    \caption{\justifying Relative OPE contributions ($\%$) for the $B_s^*$ meson as a function of $M^2$ in vacuum (a) and at $n = 5n_0, T = T_c$ (b), showing the perturbative term alongside with the three (3D-np), four (4D-np) and five dimensional (5D-np) nonperturbative contributions at $s_0 = 34\text{ GeV}^2$.}
    \label{fig:ope_contributions}
\end{figure}

The stability and reliability of the working Borel window are further corroborated through the analysis of the pole contribution. Figure~\ref{fig:pole_continuation} compares the pole contribution in vacuum and under extreme medium conditions ($T = T_c$ and $n = 5n_0$). In both cases, the pole contribution remains well above the conventional $50\%$ threshold within the designated Borel interval, confirming that the ground-state signal is not overwhelmed by the continuum background. This behavior provides an additional rigorous consistency check on the validity of the chosen Borel region, even under extreme medium modifications. To assess the convergence of the operator product expansion (OPE), we analyze the individual contributions of the perturbative and nonperturbative terms to the total sum rule.

Figure~\ref{fig:ope_contributions} shows the relative OPE contributions
for the $B_s^*$ meson in vacuum and under the extreme medium condition
$n = 5n_0$, $T = T_c$. In both cases, the perturbative term provides
the dominant contribution, followed in order by the three-dimensional nonperturbative (3D-np) operators
($\langle \bar q q\rangle$, $\langle q^\dagger q\rangle$), the four dimensional 
operators (4D-np) ($\langle\Theta_{00}^g\rangle$, $\langle q^\dagger\,iD_0
q\rangle$, $\langle\alpha_s G^2\rangle$), and the five-dimensional nonperturbative contributions  (5D-np) ($\langle\bar{q}g_s\sigma G q\rangle$, $\langle
s^\dagger\,iD_0 iD_0 s\rangle$), which remain the smallest
throughout. This hierarchy persists from vacuum to the extreme
condition, where the perturbative term still accounts for the
majority of the sum rules while the higher-dimensional condensate
contributions remain appropriately suppressed. The preservation of
this ordering ensures the rapid convergence of the OPE series within
the chosen Borel window, supporting the reliability of the QCD
sum-rule framework even under extreme thermal and dense conditions.

Before presenting the in-medium results, we quote our extracted vacuum decay 
constants in Table~\ref{tab:f0}.
\begin{table}[H]
\centering
\caption{Vacuum decay constants for $B_s^*$ and $B^*$ vector mesons.}
\label{tab:f0}
\begin{tabular}{lc}
\hline\hline
Meson(s) & $f_0$ (MeV) \\
\hline
$B_s^* \, / \, \bar{B}_s^*$             & $230.9 \pm 14.5$ \\
\hline
$B^{*\pm}$                                  & $226.6 \pm 16.0$ \\ 
$B^{*0} \, / \, \bar{B}^{*0}$               & $226.5 \pm 15.9$ \\ 
\hline\hline
\end{tabular}
\end{table}

These values serve as the baseline reference $f_0$ against which all 
finite-medium shifts are measured in Tables~\ref{tab:shifts_Bs} and \ref{tab:shifts_B}. 
The non-strange decay constants lie in a narrow range of $226$-$227~\mathrm{MeV}$, 
showing that they are insensitive to the small $u/d$ quark mass difference, 
as expected from isospin symmetry. In contrast, the strange states exhibit 
a slightly higher decay constant of approximately $231~\mathrm{MeV}$. This 
difference of about $2\%$ is due to the larger strange-quark mass and 
the corresponding condensate, which break the $SU(3)_f$ flavor symmetry.
\begin{table}[H]
\centering
\renewcommand{\arraystretch}{1.3}
\setlength{\tabcolsep}{4pt}
\caption{\justifying Relative in-medium shifts of masses and decay constants
for the $B_s^*$ multiplet under three limiting cases, evaluated as
$((\tilde{m} - m_0)/m_0)$ and $((\tilde{f} - f_0)/f_0)$.}
\label{tab:shifts_Bs}
\begin{tabular}{lcc r@{${}={}$}l r@{${}={}$}l}
\hline\hline
Meson
& $(\tilde{m} - m_0)/m_0$
& $(\tilde{f} - f_0)/f_0$
& \multicolumn{4}{c}{Condition} \\
\hline
\multirow{3}{*}{$B_s^*$}
& $-11.4\%$ & $-77.2\%$ & $T$ & $0$   & $n$ & $5n_0$ \\
& $-0.9\%$  & $-5.1\%$  & $T$ & $T_c$ & $n$ & $0$    \\
& $-11.6\%$ & $-71.5\%$ & $T$ & $T_c$ & $n$ & $5n_0$ \\
\hline
\multirow{3}{*}{$\bar{B}_s^*$}
& $-10.6\%$ & $-77.9\%$ & $T$ & $0$   & $n$ & $5n_0$ \\
& $-0.9\%$  & $-5.1\%$  & $T$ & $T_c$ & $n$ & $0$    \\
& $-11.0\%$ & $-72.0\%$ & $T$ & $T_c$ & $n$ & $5n_0$ \\
\hline\hline
\end{tabular}
\end{table}

At finite $T$ and $n$, the Borel window determined in vacuum is found to 
remain reasonably stable and is employed throughout the range up to $T \approx T_c$ 
and $n = 5n_0$. The primary parameter that varies is the effective continuum 
threshold, which decreases toward $m_b^2$ as the condensates are suppressed 
by medium effects. This behavior indicates the gradual dissolution of the 
hadronic bound states and defines the practical limit of the sum rule approach. 
Consequently, numerical results near the extreme limits of $T \approx T_c$ 
and $n \approx 5n_0$ should be interpreted with caution.
\begin{table}[H]
\centering
\renewcommand{\arraystretch}{1.3}
\setlength{\tabcolsep}{4pt}
\caption{\justifying The same as Table~\ref{tab:shifts_Bs}, but for the $B^*$ multiplet.}
\label{tab:shifts_B}
\label{tab:shifts_B}
\label{tab:shifts_B}
\begin{tabular}{lcc r@{${}={}$}l r@{${}={}$}l}
\hline\hline
Meson
& $(\tilde{m} - m_0)/m_0$
& $(\tilde{f} - f_0)/f_0$
& \multicolumn{4}{c}{Condition} \\
\hline

\multirow{3}{*}{$B^{*+}$}
& $-8.3\%$  & $-79.6\%$ & $T$ & $0$   & $n$ & $5n_0$ \\
& $-0.6\%$  & $-4.8\%$  & $T$ & $T_c$ & $n$ & $0$    \\
& $-8.4\%$  & $-74.3\%$ & $T$ & $T_c$ & $n$ & $5n_0$ \\
\hline

\multirow{3}{*}{$B^{*-}$}
& $-12.6\%$ & $-86.2\%$ & $T$ & $0$   & $n$ & $5n_0$ \\
& $-1.1\%$  & $-5.3\%$  & $T$ & $T_c$ & $n$ & $0$    \\
& $-12.8\%$ & $-78.3\%$ & $T$ & $T_c$ & $n$ & $5n_0$ \\
\hline

\multirow{3}{*}{$B^{*0}$}
& $-6.1\%$  & $-79.1\%$ & $T$ & $0$   & $n$ & $5n_0$ \\
& $-0.5\%$  & $-4.9\%$  & $T$ & $T_c$ & $n$ & $0$    \\
& $-6.8\%$  & $-74.1\%$ & $T$ & $T_c$ & $n$ & $5n_0$ \\
\hline

\multirow{3}{*}{$\bar{B}^{*0}$}
& $-12.9\%$ & $-69.0\%$ & $T$ & $0$   & $n$ & $5n_0$ \\
& $-1.1\%$  & $-3.9\%$  & $T$ & $T_c$ & $n$ & $0$    \\
& $-12.3\%$ & $-63.7\%$ & $T$ & $T_c$ & $n$ & $5n_0$ \\
\hline\hline
\end{tabular}
\end{table}
The non-strange and strange sectors differ mainly in the magnitude of the 
induced particle-antiparticle asymmetry. In the non-strange sector, this 
asymmetry increases significantly with density, whereas it remains small 
in the strange sector across all investigated densities and temperatures. The origin of this difference lies in the strange-quark sector, which we analyze in Sec. III e.

\begin{figure}[H]
\centering
\hspace*{-0.1\columnwidth}
\includegraphics[width=1.05\columnwidth]{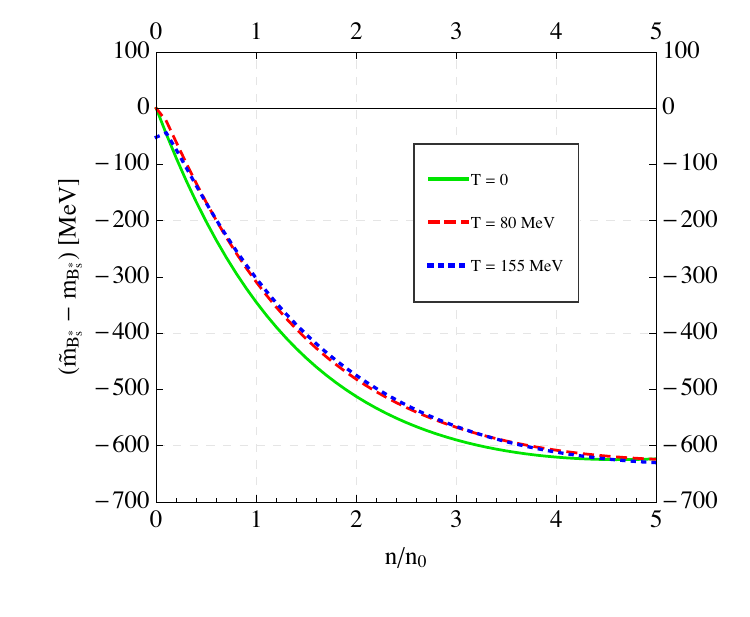}
\caption{\justifying In-medium mass shift of the $B_s^*$ meson, 
$ \tilde{m}_{B_s^*} - m_{B_s^*}$ (in MeV), as a function of the 
normalized baryon density $n/n_0$ at fixed temperatures $T = 0$, $80$, and $155~\mathrm{MeV}$.}
\label{fig:BsStarMassDensity}
\end{figure}

To give a clearer picture of the medium effects, Tables~\ref{tab:shifts_Bs} and
\ref{tab:shifts_B} present the relative changes in the masses and decay constants
of the $B_s^*$ and $B^*$ multiplets for three representative temperature and
density conditions. Temperature alone (at $n=0$) produces
only modest shifts, of at most $1.1\%$ for the masses and $(3.9$-$5.3)\%$ for
the decay constants, whereas the combined effect of high temperature and density
leads to substantial suppressions, particularly in the non-strange sector.

\begin{figure}[H]
\centering
\hspace*{-0.1\columnwidth}
\includegraphics[width=1.05\columnwidth]{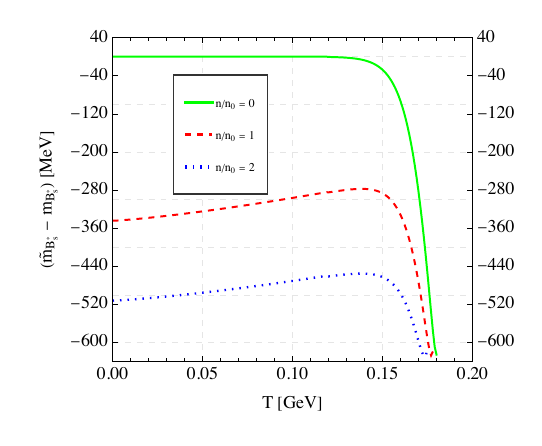}
\caption{\justifying The same mass shift as in Fig.~\ref{fig:BsStarMassDensity}, 
but plotted as a function of temperature $T$ at fixed densities 
$n/n_0 = 0$, $1$, and $2$.}
\label{fig:BsStarMassT}
\end{figure}

\paragraph{\textbf{$B_s^*$ meson mass}}
Figs.~\ref{fig:BsStarMassDensity} and~\ref{fig:BsStarMassT} show how the $B_s^*$ mass shifts with density and temperature. The drop is monotonic in both directions. At high density, the three temperature curves nearly overlap, indicating that temperature is a secondary factor relative to density. Temperature has almost no effect on the mass below $\sim 0.14$~GeV; the sharp suppression occurs only in the vicinity of $T_c$. To analyze these medium-induced modifications more quantitatively, we focus on the specific physical regimes highlighted in our previous tables:

\begin{itemize}
\item \textbf{Cold dense matter ($T = 0,\ n = 5n_0$):}
As Table~\ref{tab:shifts_Bs} shows, at zero temperature and
high baryon density the $B_s^{*}$ mass decreases slightly more
than that of its antiparticle: $-11.4\%$ against $-10.6\%$. In
cold dense matter, $B_s^{*}$ is therefore lighter than
$\bar{B}_s^{*}$, though the gap stays below one percent of the
vacuum mass.

\item \textbf{Pure temperature ($T = T_c,\ n = 0$):}
The mass is almost completely insensitive to temperature, with
a shift of less than $1\%$ at $T_c$. This reflects the
protection provided by the heavy $b$-quark, whose large bare
mass limits how far the in-medium condensate suppression can
reduce the meson mass. With no net baryon charge in the medium,
particle and antiparticle respond identically.
\begin{figure}[H]
\centering
\hspace*{-0.1\columnwidth}
\includegraphics[width=1.05\columnwidth]{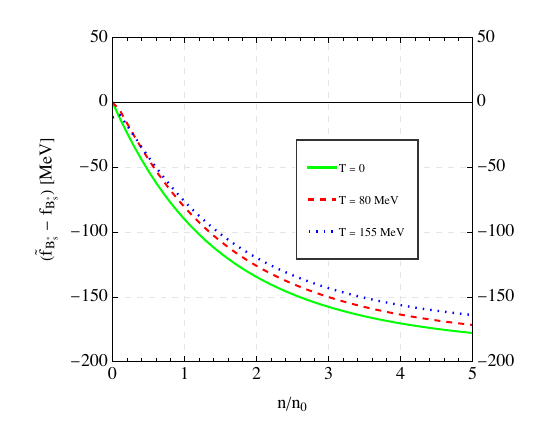}
\caption{\justifying In-medium shift of the $B_s^*$ decay constant, 
$ \tilde{f}_{B_s^*} - f_{B_s^*}$ (in MeV), as a function of the 
normalized baryon density $n/n_0$ at fixed temperatures $T = 0$, $80$, and $155~\mathrm{MeV}$.}
\label{fig:BsStarfDensity}
\label{fig:BsStarfDensity}
\end{figure}
\item \textbf{Combined effects ($T = T_c,\ n = 5n_0$):}
Heating changes little. The shifts move to $-11.6\%$ for
$B_s^{*}$ and $-11.0\%$ for $\bar{B}_s^{*}$, so the small
particle-antiparticle gap of the cold dense case survives in
slightly reduced form from about $42$~MeV at $T=0$ to about
$31$~MeV at $T_c$. The asymmetry in the strange sector is thus
a mild but persistent density effect.
\end{itemize}

\paragraph{\textbf{$B_s^*$ decay constant}}
\begin{figure}[H]
\centering
\hspace*{-0.1\columnwidth}
\includegraphics[width=1.05\columnwidth]{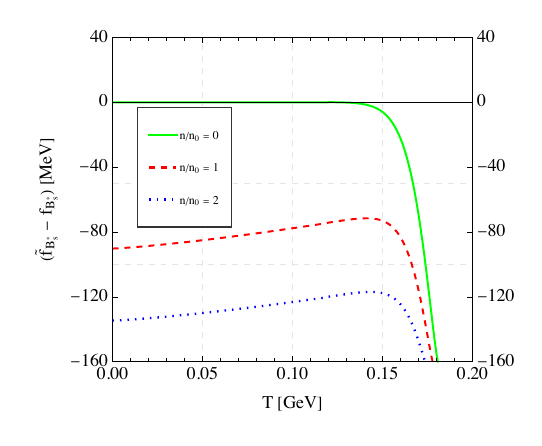}
\caption{\justifying The same decay constant shift as in Fig.~\ref{fig:BsStarfDensity}, 
but plotted as a function of temperature $T$ at fixed densities 
$n/n_0 = 0$, $1$, and $2$.}
\label{fig:BsStarfT}
\label{fig:BsStarfT}
\end{figure}
The decay constants, shown in Figs.~\ref{fig:BsStarfDensity} and
\ref{fig:BsStarfT}, respond far more strongly to the medium than the
masses: the large $b$-quark mass limits the extent to which the
hadron mass can shift, whereas the decay constant directly reflects
the reduction of the QCD condensates.  Table~\ref{tab:shifts_Bs} shows that baryon density is the primary driver of in-medium modifications for the $B_s^*$ multiplet, drastically suppressing decay constants by up to $\sim 78\%$ and masses by $\sim 11\%$ at $5n_0$. To examine these pronounced variations in more detail, we now evaluate the 
decay constant shifts under the same representative limiting cases:
\begin{itemize}
\item \textbf{Cold dense regime ($T = 0,\ n = 5n_0$):}
In contrast to the pure thermal case, high baryon density
drives a much stronger suppression of the decay constants,
reducing them by $\sim 77$-$78\%$. A possible interpretation
is that the dense nuclear matter screens the color fields
between the quarks, weakening their binding and spreading the
pair apart. The decay constant, which essentially measures how
often the two quarks overlap at a single point, would then fall
accordingly.

\item \textbf{Pure thermal regime ($T = T_c,\ n = 0$):}
At vanishing baryon density, raising the temperature to $T_c$
reduces the decay constants by only about $5\%$, and by the
same amount for both states. In the absence of a net baryon density, the medium preserves the 
particle-antiparticle degeneracy. Consequently, no asymmetry is induced, 
and the decay constants of the $B_s^{*}$ and $\bar{B}_s^{*}$ states 
remain identical at $n=0$, reflecting the symmetric vacuum baseline presented 
in Table~\ref{tab:f0}.

\item \textbf{Combined hot and dense regime ($T = T_c,\ n = 5n_0$):}
Under the simultaneous influence of finite temperature and density, the suppression 
of the $B_s^{*}$ decay constant ($-71.5\%$) is slightly less pronounced than 
in the cold dense case ($-77.2\%$), with a similar trend observed for $\bar{B}_s^{*}$. 
Within the QCD sum rule framework, this behavior arises from a partial cancellation 
between the temperature- and density-dependent OPE contributions, which moderates 
the overall in-medium reduction.\end{itemize}

\begin{figure}[H]
\centering
\hspace*{-0.1\columnwidth}
\includegraphics[width=0.5\textwidth,keepaspectratio]%
  {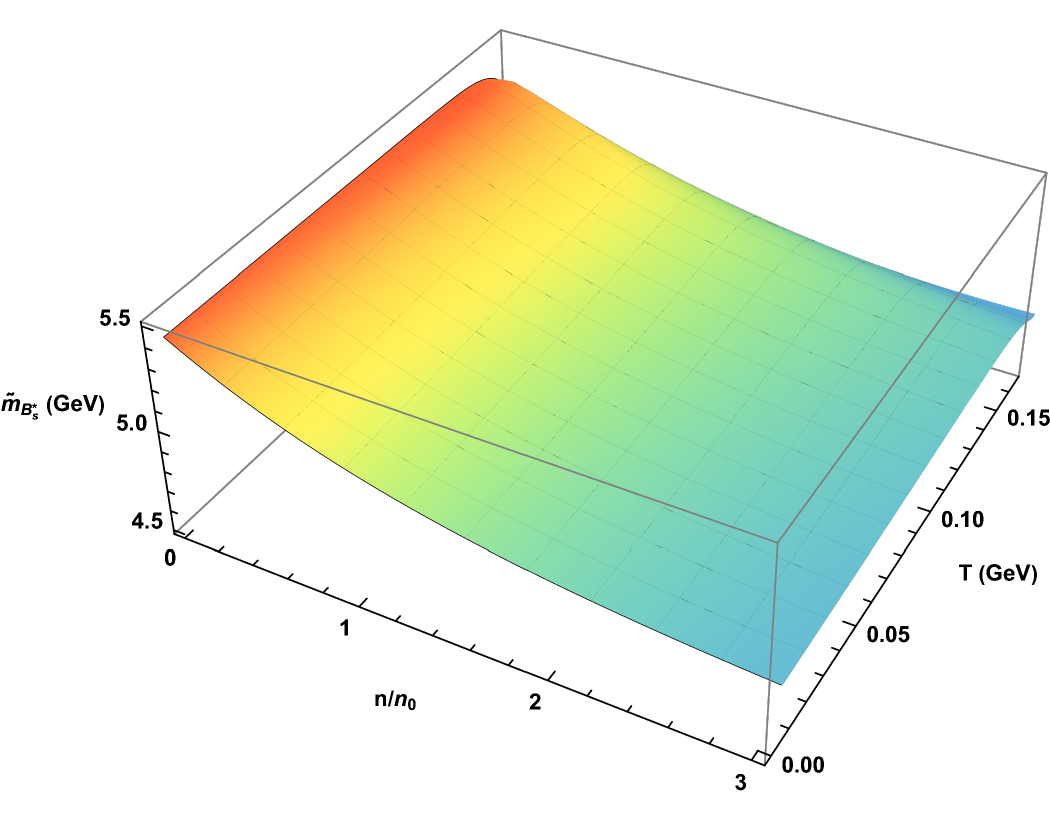}
\caption{\justifying The in-medium mass $\tilde{m}_{B_s^*}(T,n)$ (in GeV) of the 
$B_s^*$ meson, plotted as a function of temperature $T$ and baryon density $n/n_0$.}
\label{fig:BsStarMass3Da}
\end{figure}

\begin{figure}[H]
\centering
\hspace*{-0.1\columnwidth}
\includegraphics[width=0.5\textwidth,keepaspectratio]%
  {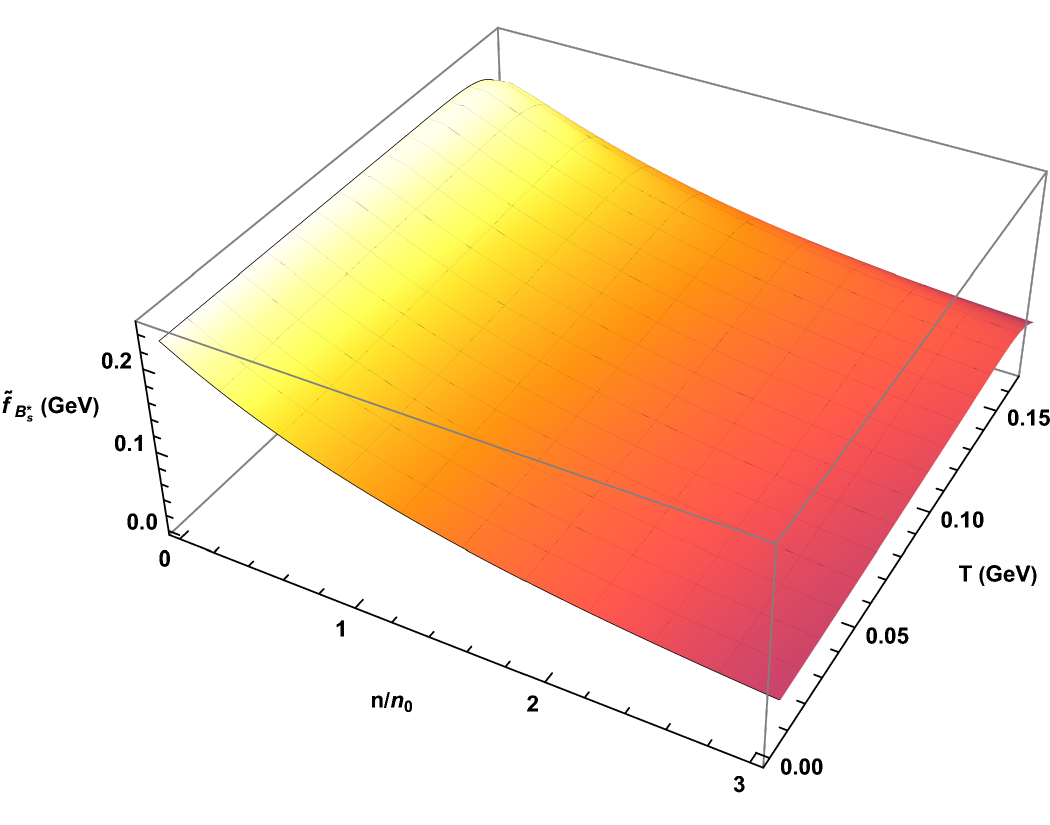}
\caption{\justifying The in-medium decay constant $\tilde{f}_{B_s^*}(T,n)$ (in GeV) 
of the $B_s^*$ meson, plotted as a function of temperature $T$ and baryon density $n/n_0$.}
\label{fig:BsStarf3Db}
\end{figure}

\begin{figure}[H]
\centering
\begin{subfigure}[b]{1.05\columnwidth}
\hspace*{-0.1\columnwidth}
  \centering
  \includegraphics[width=\textwidth,height=7.5cm,keepaspectratio]%
    {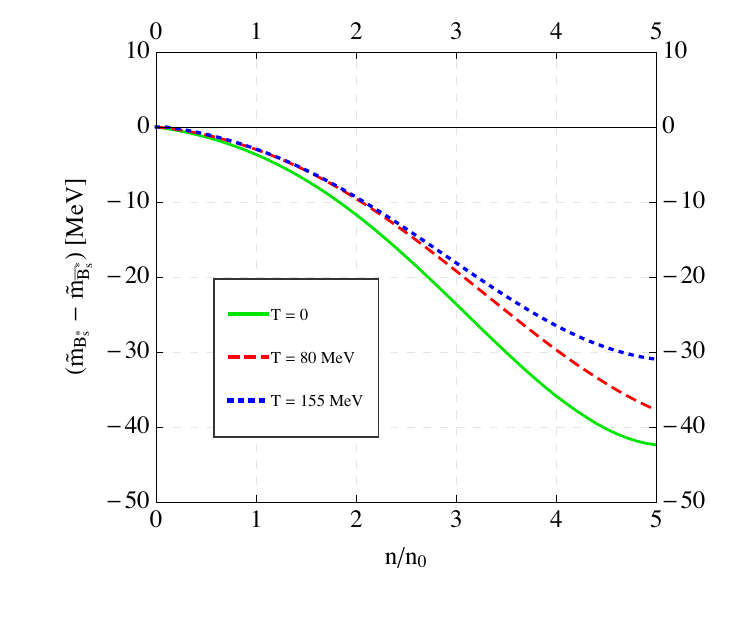}
  \caption{$(\tilde{m}_{B_s^*} - \tilde{m}_{\bar{B}_s^*})$~[MeV]
    vs $n/n_0$.}
\end{subfigure}

\vspace{4pt}

\begin{subfigure}[b]{1.05\columnwidth}
  \centering
  \hspace*{-0.1\columnwidth}
  \includegraphics[width=\textwidth,height=7.5cm,keepaspectratio]%
    {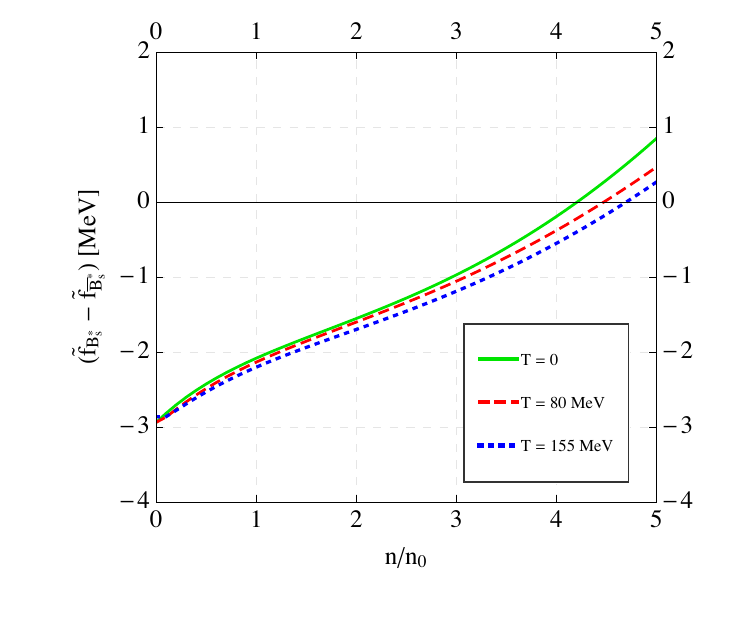}
  \caption{$(\tilde{f}_{B_s^*} - \tilde{f}_{\bar{B}_s^*})$~[MeV]
    vs $n/n_0$.}
\end{subfigure}
\caption{\justifying In-medium $B_s^*$-$\bar{B}_s^*$ mass (top) and decay-constant 
(bottom) splittings as functions of baryon density at fixed $T=0$, $80$, and $155~\mathrm{MeV}$.}
\label{fig:BsStarSplitDensity}
\end{figure}

\begin{figure}[H]
\centering
\begin{subfigure}[b]{1\columnwidth}
\hspace*{-0.1\columnwidth}
  \centering
  \includegraphics[width=\textwidth,height=7.5cm,keepaspectratio]%
    {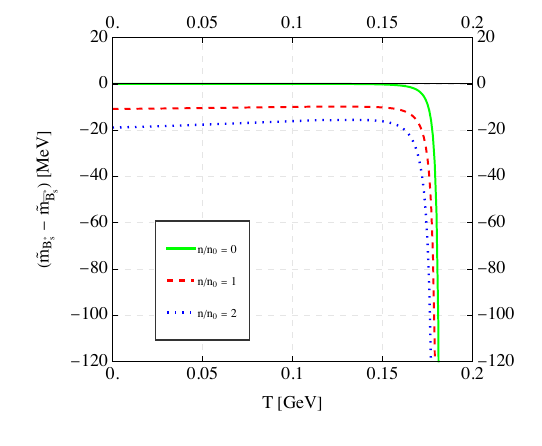}
  \caption{$(\tilde{m}_{B_s^*} - \tilde{m}_{\bar{B}_s^*})$ [MeV]
    vs $T$.}
\end{subfigure}

\vspace{15pt}

\begin{subfigure}[b]{1\columnwidth}
\hspace*{-0.1\columnwidth}
  \centering
  \includegraphics[width=\textwidth,height=7.5cm,keepaspectratio]%
    {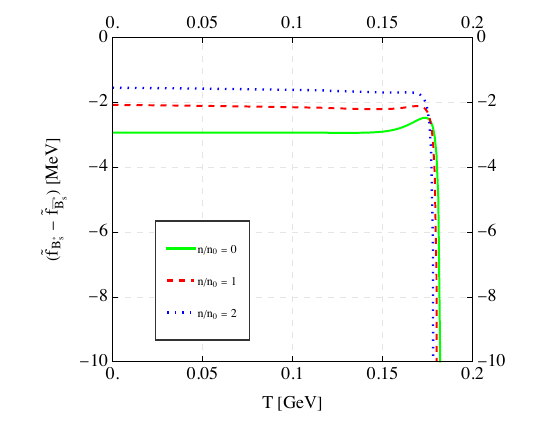}
  \caption{$(\tilde{f}_{B_s^*} - \tilde{f}_{\bar{B}_s^*})$ [MeV]
    vs $T$.}
\end{subfigure}
\caption{\justifying In-medium $B_s^*$-$\bar{B}_s^*$ mass (top) and decay-constant 
(bottom) splittings as functions of temperature $T$ at fixed density ratios 
$n/n_0 = 0$, $1$, and $2$.}
\label{fig:BsStarSplitT}
\end{figure}

\begin{figure}[H]
\centering
\begin{subfigure}[b]{1\columnwidth}
\hspace*{-0.1\columnwidth}
  \centering
  \includegraphics[width=\textwidth,height=7.5cm,keepaspectratio]%
    {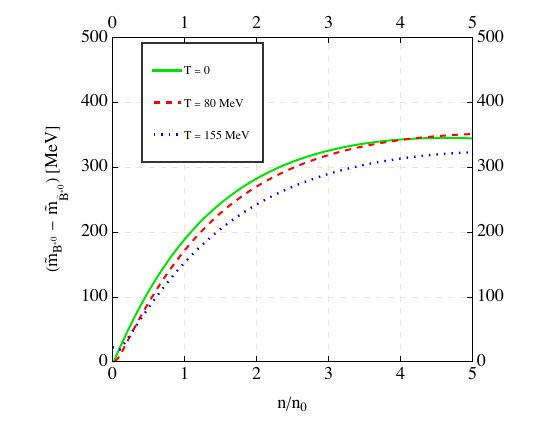}
  \caption{$(\tilde{m}_{B^{*0}} - \tilde{m}_{\bar{B}^{*0}})$~[MeV]
    vs $n/n_0$.}
\end{subfigure}

\vspace{4pt}

\begin{subfigure}[b]{1\columnwidth}
\hspace*{-0.1\columnwidth}
  \centering
  \includegraphics[width=\textwidth,height=7.5cm,keepaspectratio]%
    {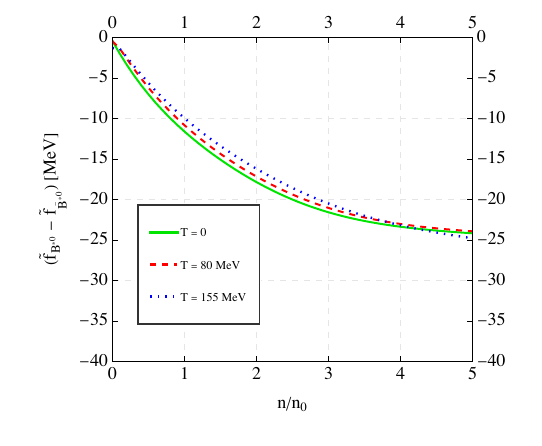}
  \caption{$(\tilde{f}_{B^{*0}} - \tilde{f}_{\bar{B}^{*0}})$~[MeV]
    vs $n/n_0$.}
\end{subfigure}
\caption{\justifying In-medium $B^{*0}$-$\bar{B}^{*0}$ mass (top) and decay-constant 
(bottom) splittings as functions of baryon density at $T=0$, $80$, and $155~\mathrm{MeV}$.}
\label{fig:BStarSplitsNeutral}
\end{figure}

\begin{figure}[H]
\centering
\begin{subfigure}[b]{0.95\columnwidth}
\hspace*{-0.1\columnwidth}
  \centering
  \includegraphics[width=\textwidth,height=7.5cm,keepaspectratio]%
    {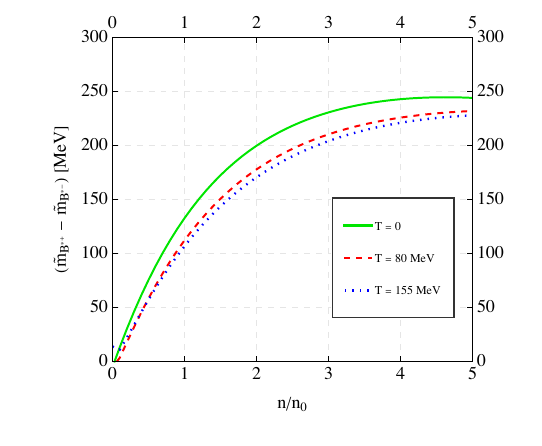}
  \caption{$(\tilde{m}_{B^{*+}} - \tilde{m}_{B^{*-}})$~[MeV]
    vs $n/n_0$.}
\end{subfigure}

\vspace{4pt}

\begin{subfigure}[b]{0.95\columnwidth}
  \centering
  \hspace*{-0.1\columnwidth}
  \includegraphics[width=\textwidth,height=7.5cm,keepaspectratio]%
    {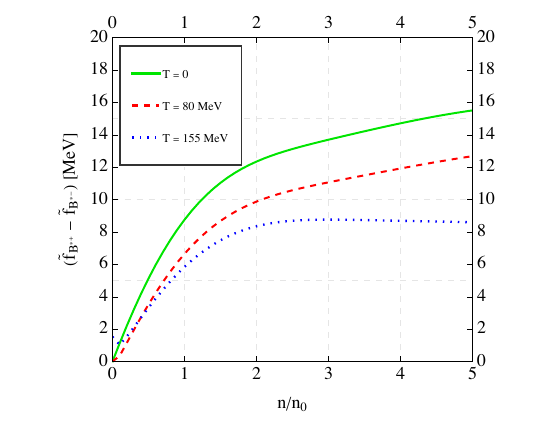}
  \caption{$(\tilde{f}_{B^{*+}} - \tilde{f}_{B^{*-}})$~[MeV]
    vs $n/n_0$.}
\end{subfigure}
\caption{\justifying In-medium $B^{*+}$-$B^{*-}$ mass (top) and decay-constant 
(bottom) splittings as functions of baryon density at $T=0$, $80$, and $155~\mathrm{MeV}$.}
\label{fig:BStarSplitsCharged}
\end{figure}

\FloatBarrier
\paragraph{\textbf{Three-dimensional view}}
The surfaces in Figs.~\ref{fig:BsStarMass3Da} and~\ref{fig:BsStarf3Db} illustrate 
the distinct characteristics of the two in-medium effects. Along the density axis, 
the reduction in both parameters initiates immediately and progresses monotonically. 
In contrast, along the temperature axis, the spectroscopic properties remain nearly 
unaffected over a wide range, staying flat up to $T \sim 0.14~\mathrm{GeV}$ before 
exhibiting a rapid suppression in the vicinity of $T_c$. Consequently, the most 
pronounced medium-induced modifications occur in the high-temperature, high-density 
regime ($T \approx T_c$ and $n \approx 5n_0$), which also marks the limit of 
the reliability of our sum-rule calculation.
%
\paragraph{\textbf{$B^*$ multiplet}}
Table~\ref{tab:shifts_B} summarizes the in-medium parameters for the $B^*$ multiplet 
under the three limiting cases, while Fig.~\ref{fig:BStarShifts} displays the continuous 
density dependence of the mass and decay-constant shifts for all four non-strange states. 
Unlike the $B_s^*$ sector, where particle and antiparticle states respond almost symmetrically, 
the non-strange sector exhibits a prominent asymmetry that strongly depends on the specific 
physical regime. To analyze this state-dependent behavior quantitatively, we focus on 
the individual modifications under the key physical conditions:
\begin{figure*}[tp]
\centering
\begin{subfigure}[t]{0.50\textwidth}
\hspace*{-0.1\columnwidth}
  \includegraphics[width=\textwidth,height=5.3cm,keepaspectratio]%
    {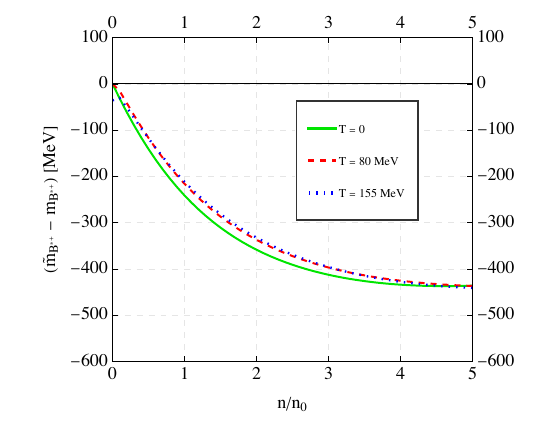}
  \caption{$(\tilde{m}_{B^{*+}} - m_{B^{*+}})$~[MeV] vs $n/n_0$.}
\end{subfigure}\hfill
\begin{subfigure}[t]{0.50\textwidth}
\hspace*{-0.1\columnwidth}
  \includegraphics[width=\textwidth,height=5.3cm,keepaspectratio]%
    {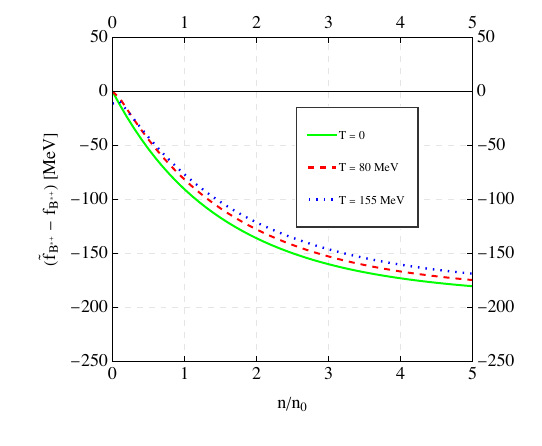}
  \caption{$(\tilde{f}_{B^{*+}} - f_{B^{*+}})$~[MeV] vs $n/n_0$.}
\end{subfigure}

\vspace{1pt}

\begin{subfigure}[t]{0.50\textwidth}
\hspace*{-0.1\columnwidth}
  \includegraphics[width=\textwidth,height=5.3cm,keepaspectratio]%
    {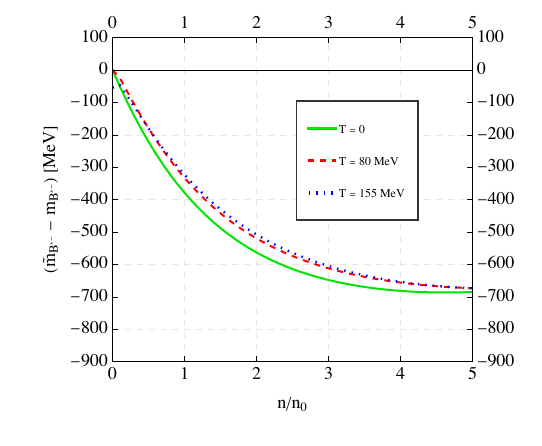}
  \caption{$(\tilde{m}_{B^{*-}} - m_{B^{*-}})$~[MeV] vs $n/n_0$.}
\end{subfigure}\hfill
\begin{subfigure}[t]{0.50\textwidth}
\hspace*{-0.1\columnwidth}
  \includegraphics[width=\textwidth,height=5.3cm,keepaspectratio]%
    {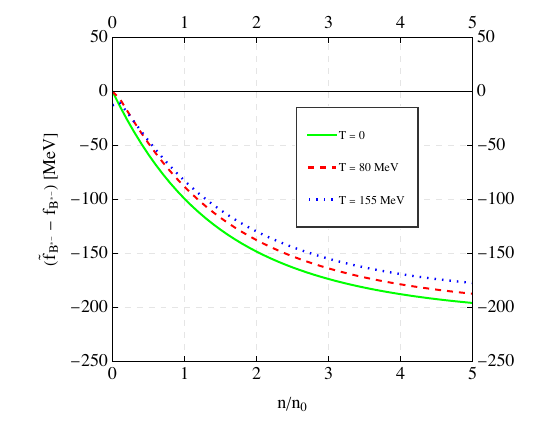}
  \caption{$(\tilde{f}_{B^{*-}} - f_{B^{*-}})$~[MeV] vs $n/n_0$.}
\end{subfigure}

\vspace{1pt}

\begin{subfigure}[t]{0.50\textwidth}
\hspace*{-0.1\columnwidth}
  \includegraphics[width=\textwidth,height=5.3cm,keepaspectratio]%
    {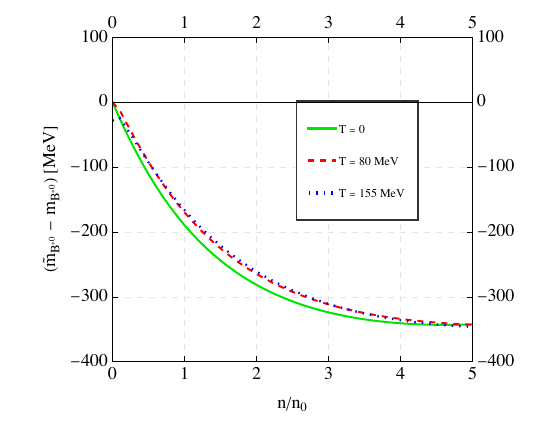}
  \caption{$(\tilde{m}_{B^{*0}} - m_{B^{*0}})$~[MeV] vs $n/n_0$.}
\end{subfigure}\hfill
\begin{subfigure}[t]{0.50\textwidth}
\hspace*{-0.1\columnwidth}
  \includegraphics[width=\textwidth,height=5.3cm,keepaspectratio]%
    {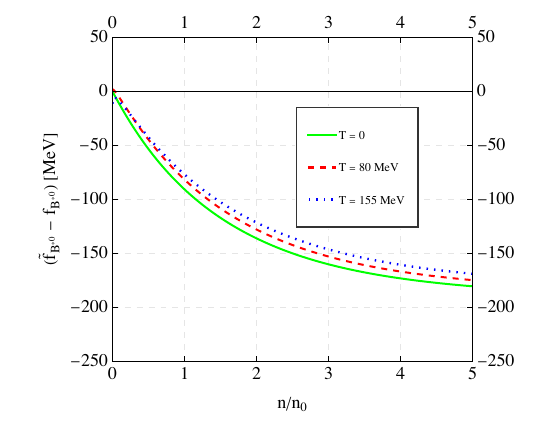}
  \caption{$(\tilde{f}_{B^{*0}} - f_{B^{*0}})$~[MeV] vs $n/n_0$.}
\end{subfigure}

\vspace{1pt}

\begin{subfigure}[t]{0.50\textwidth}
\hspace*{-0.1\columnwidth}
  \includegraphics[width=\textwidth,height=5.3cm,keepaspectratio]%
    {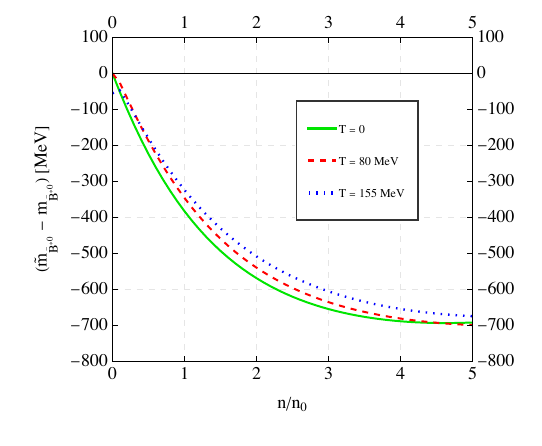}
  \caption{$(\tilde{m}_{\bar{B}^{*0}} - m_{\bar{B}^{*0}})$~[MeV] vs $n/n_0$.}
\end{subfigure}\hfill
\begin{subfigure}[t]{0.50\textwidth}
\hspace*{-0.1\columnwidth}
  \includegraphics[width=\textwidth,height=5.3cm,keepaspectratio]%
    {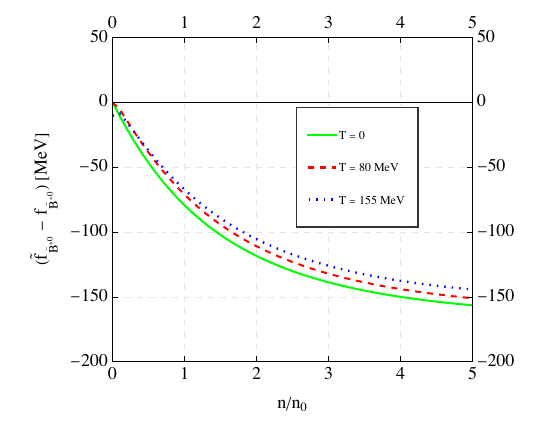}
  \caption{$(\tilde{f}_{\bar{B}^{*0}} - f_{\bar{B}^{*0}})$~[MeV] vs $n/n_0$.}
\end{subfigure}
\caption{\justifying In-medium mass (left) and decay-constant
(right) shifts of $B^{*+}$, $B^{*-}$, $B^{*0}$, and $\bar{B}^{*0}$
vs baryon density at $T=0$, $80$~MeV,
$155$~MeV.}
\label{fig:BStarShifts}
\end{figure*}
\begin{itemize}
    \item \textbf{Cold dense case ($T = 0$, $n = 5n_0$):} High baryon density induces the 
    most substantial suppressions, though the four states exhibit a non-uniform response. 
    The decay constants experience the most pronounced reduction for $B^{*-}$ ($-86.2\%$), 
    followed by $B^{*+}$ ($-79.6\%$) and $B^{*0}$ ($-79.1\%$), whereas $\bar{B}^{*0}$ is 
    the least affected ($-69.0\%$). A similar asymmetry manifests in the mass shifts: 
    the masses of $B^{*-}$ ($-12.6\%$) and $\bar{B}^{*0}$ ($-12.9\%$) decrease the most, 
    while those of $B^{*+}$ and $B^{*0}$ decrease by only $-8.3\%$ and $-6.1\%$, respectively. 
    This mass splitting is driven by the vector self-energy $\Sigma_v$, which enters with 
    opposite signs for particles and antiparticles, thereby leading to a larger mass 
    reduction in the antiparticle states ($B^{*-}$, $\bar{B}^{*0}$) compared to their 
    particle counterparts ($B^{*+}$, $B^{*0}$). Remarkably, the decay constants do not 
    follow this ordering; despite experiencing the largest mass shift, $\bar{B}^{*0}$ retains 
    its vacuum decay constant more effectively than the other states. This divergence 
    indicates that the in-medium modifications of masses and decay constants probe 
    distinct aspects of the nuclear medium.

\item \textbf{Pure thermal case ($T = T_c$, $n = 0$):} In the absence of 
baryon density, all four states exhibit closely aligned in-medium modifications, 
with the decay constants decreasing by $3.9\%$-$5.3\%$ and the masses by 
$0.5\%$-$1.1\%$. Since the net baryon density vanishes, the medium 
preserves the particle-antiparticle symmetry, while the large mass of the heavy 
$b$-quark fundamentally limits the magnitude of the thermal modifications. 
Consequently, the resulting suppression remains marginal compared to the 
density-dominated regimes.

\item \textbf{Combined hot and dense case ($T = T_c$, $n = 5n_0$):} Under the 
simultaneous influence of finite temperature and density, the particle-antiparticle 
asymmetry observed in the cold dense regime is partially suppressed but remains 
manifest. The in-medium mass shifts range from $-6.8\%$ for $B^{*0}$ to $-12.8\%$ 
for $B^{*-}$, while the decay constant reductions span $63.7\%$-$78.3\%$, with 
$\bar{B}^{*0}$ showing the smallest suppression ($-63.7\%$) and $B^{*-}$ the largest 
($-78.3\%$). Thermal fluctuations in the vicinity of $T_c$ slightly moderate the 
$\Sigma_v$-driven mass splitting; nevertheless, the masses of $B^{*0}$ and $B^{*+}$ 
remain distinctly less suppressed than those of $B^{*-}$ and $\bar{B}^{*0}$.
\end{itemize}
\paragraph{\textbf{Particle-antiparticle splittings}}
Figs.~\ref{fig:BsStarSplitDensity}-\ref{fig:BStarSplitsCharged} show the medium-induced 
splittings of the multiplet as functions of baryon density and temperature, expressed as the mass 
and decay-constant differences between pairs of states, $\tilde{m}_X - \tilde{m}_Y$ 
and $\tilde{f}_X - \tilde{f}_Y$. Here, $X$ and $Y$ denote the two states under comparison: 
a particle and its antiparticle in the present discussion, or two states with different accompanying 
light-quark flavors in the isospin analysis below. Following the PDG
classification~\cite{PDG:2024}, $B^{*+}\,(\bar{b}u)$ and
$B^{*0}\,(\bar{b}d)$ are the particle states of the multiplet,
and $B^{*-}\,(b\bar{u})$ and $\bar{B}^{*0}\,(b\bar{d})$ are the
antiparticle states. 

In the vacuum limit, all splittings strictly vanish, ensuring that any 
particle-antiparticle difference observed in these profiles is genuinely 
generated by the medium. The well-established dynamics of the kaon system 
provide a useful baseline for this behavior: in nuclear matter, the WT
vector interaction enters with opposite signs for states containing a light quark, 
such as $K^+\,(u\bar{s})$, versus those containing a light antiquark, 
such as $K^-\,(\bar{u}s)$~\cite{Weinberg:1966kf,Tomozawa:1966jm}, thereby driving 
the $K$ and $\bar{K}$ masses apart with increasing density. An analogous mechanism 
governs the $B^*$ multiplet: the presence of light quarks in $B^{*+}$ and $B^{*0}$ 
versus light antiquarks in $B^{*-}$ and $\bar{B}^{*0}$ causes the vector self-energy 
$\Sigma_v$ to contribute with opposite signs for the two subgroups, ultimately 
leading to the observed asymmetry in-medium modifications~\cite{Cohen:1994wm}.
The strange pair exhibits the most constrained splittings among the investigated multiplets 
(Figs.~\ref{fig:BsStarSplitDensity} and~\ref{fig:BsStarSplitT}), with the maximum in-medium variations reaching
\vspace{15pt}
\begin{align}
 \tilde{m}_{B_s^*} - \tilde{m}_{\bar{B}_s^*} \approx -42.5~\mathrm{MeV}, \nonumber\\
 \tilde{f}_{B_s^*} - \tilde{f}_{\bar{B}_s^*} \approx +0.8~\mathrm{MeV},
\label{eq:split_strange}
\end{align}
at $n = 5n_0$ and $T=0$. Across the entire phase space, the decay-constant splitting 
displays remarkable resilience, varying by less than $4~\mathrm{MeV}$. Conversely, the mass splitting 
 develops progressively with density—reaching its peak magnitude in the cold, dense limit 
($T=0, n=5n_0$)—before undergoing thermal suppression as $T \to T_c$.
%

This suppressed response in the strange sector originates directly from the hadronic composition of the medium. 
Since nucleons contain no valence strange quarks, the direct vector density operator vanishes for strangeness, 
$\langle s^\dagger s\rangle_{T,n} = 0$, in contrast to the light-quark sector where $\langle q^\dagger q\rangle_{T,n} = \tfrac{3}{2}n$. Consequently, the strange condensate feels the nuclear medium only indirectly through the sub-leading 
flavor-mixing relation $\langle\bar{s}s\rangle(T,n) \approx 0.8\,\langle\bar{q}q\rangle(T,n)$, thereby moderating 
the overall magnitude of the in-medium splitting.

It should be noted that while CPT invariance strictly enforces identical vacuum decay constants for particles  and antiparticles [$f_0(B_s^*) = f_0(\bar{B}_s^*)$], as reported in Table~\ref{tab:f0}, the raw numerical evaluations in the parameter plots retain a minor vacuum baseline offset of approximately $-3~\mathrm{MeV}$. This small numerical artifact arises from the independent sum-rule optimizations performed for the individual channels. 
However, for the physical interpretation of the in-medium effects, all decay-constant shifts discussed herein are evaluated relative to this baseline, ensuring that the physical conclusions remain strictly invariant under  the choice of the vacuum reference. This pronounced response in the $\bar{b}d$/$b\bar{d}$ sector arises because both the $u$- and $d$-quark 
dense backgrounds contribute constructively to the vector self-energy $\Sigma_v$. By contrast, in the charged 
pair, only the $u$-quark density contributes directly. The charged states thus follow a qualitatively 
similar pattern, albeit with reduced overall magnitudes (Fig.~\ref{fig:BStarSplitsCharged}):
\begin{align}
 \tilde{m}_{B^{*+}} - \tilde{m}_{B^{*-}} \approx +240~\mathrm{MeV}, \nonumber\\
 \tilde{f}_{B^{*+}} - \tilde{f}_{B^{*-}} \approx +16~\mathrm{MeV}.
\label{eq:split_charged}
\end{align}

Thermal modifications leave both observables in the charged sector largely unaffected up to $T \to T_c$: 
the mass splitting retains a high value of approximately $+230~\mathrm{MeV}$, whereas the decay-constant 
splitting decreases to roughly half its cold-medium value, stabilizing near $+9~\mathrm{MeV}$.

Although the presence of both light-quark flavors in the background
qualitatively explains why the neutral pair experiences a stronger
medium modification than the charged pair, a fully quantitative
decomposition of this asymmetry would require a detailed, term-by-term
breakdown of the vector self-energy $\Sigma_v$, which we defer to
future investigation. Nonetheless, it is worth commenting briefly on
the overall magnitude of the resulting neutral-pair mass splitting,
quoted below. It should be stressed that this splitting is not simply
$2\Sigma_v$; the self-energy enters hadronic side of the sum rule, and
the extracted masses reflect its combined effect on the pole position.
Moreover, $n=5n_0$ is an extreme density chosen to bound the maximal
medium effect: at normal nuclear density $n_0$ the splitting is
considerably smaller, about $188.3$~MeV, so the results near
$n \approx 5n_0$ should be viewed as an upper estimate rather than a
precise prediction.

This in-medium asymmetric behavior is displayed most prominently by the neutral pair (Fig.~\ref{fig:BStarSplitsNeutral}), 
yielding the largest splittings across the entire multiplet:
\begin{align}
 \tilde{m}_{B^{*0}} - \tilde{m}_{\bar{B}^{*0}} \approx +350~\mathrm{MeV}, \nonumber\\
 \tilde{f}_{B^{*0}} - \tilde{f}_{\bar{B}^{*0}} \approx -24~\mathrm{MeV},
\label{eq:split_neutral}
\end{align}
at $n=5n_0$ and $T=0$. Notably, the two observables evolve in opposite directions with increasing density. 
The mass splitting grows systematically with density, while thermal effects induce a moderate suppression, 
reducing the gap to approximately $+300~\mathrm{MeV}$ near $T_c$. Conversely, the decay-constant 
splitting deepens with density and displays remarkable thermal resilience, persisting in the range of 
$-(20\text{-}24)~\mathrm{MeV}$ up to $T_c$. Consequently, the neutral pair retains a substantial fraction 
of its decay-constant asymmetry even in the high-temperature regime, whereas its mass asymmetry is 
attenuated but not erased. 

Taken together, these three multiplets demonstrate that the in-medium mass and 
decay-constant splittings respond to thermal effects through qualitatively 
distinct mechanisms. The mass splittings, which scale primarily with baryon 
density, undergo gradual thermal suppression and tend toward vanishing as $T \to T_c$, 
confirming that this asymmetry is predominantly a density-driven phenomenon 
rather than a thermal one. In contrast, the decay-constant splittings display a markedly 
non-uniform behavior across the flavors: the neutral splitting is partially preserved 
under heating, the charged one is substantially reduced, and the strange splitting 
remains negligible across all temperatures. This latter resilience reflects the 
relative stability of the strange quark condensate, which exhibits only marginal 
temperature dependence in this domain.

\paragraph{\textbf{Isospin splitting}}
A second, independent comparison keeps the heavy-quark content fixed and 
changes only the light-quark flavor ($d$ versus $u$). Since $B^{*0}$ ($\bar{b}d$) 
and $B^{*+}$ ($\bar{b}u$) are both particle states, this isospin splitting isolates 
the light-flavor asymmetric response of the nuclear medium without direct contributions 
from the vector self-energy $\Sigma_v$.

\begin{figure}[H]
\centering
\begin{subfigure}[b]{1\columnwidth}
\hspace*{-0.1\columnwidth}
  \centering
  \includegraphics[width=\textwidth,height=7.5cm,keepaspectratio]%
    {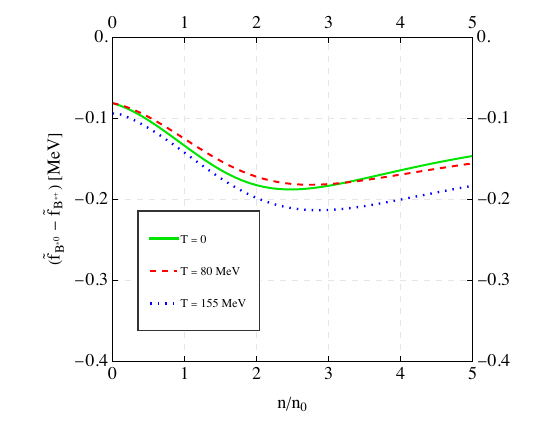}
  \caption{$(\tilde{f}_{B^{*0}} - \tilde{f}_{B^{*+}})$ [MeV] vs $n/n_0$.}
\end{subfigure}

\vspace{4pt}

\begin{subfigure}[b]{1\columnwidth}
\hspace*{-0.1\columnwidth}
  \centering
  \includegraphics[width=\textwidth,height=7.5cm,keepaspectratio]%
    {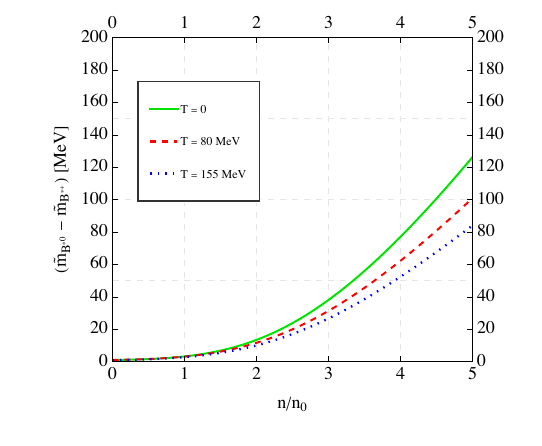}
  \caption{$(\tilde{m}_{B^{*0}} - \tilde{m}_{B^{*+}})$ [MeV] vs $n/n_0$.}
\end{subfigure}
\caption{\justifying In-medium $B^{*0}$-$B^{*+}$ isospin (a) decay-constant and 
(b) mass splittings as functions of baryon density at $T=0$, $80$, and $155~\mathrm{MeV}$.}
\label{fig:BStarSplitsIsospin}
\end{figure}

Fig.~\ref{fig:BStarSplitsIsospin} illustrates the density and thermal evolution 
of these differences. At the maximum evaluated density ($n = 5n_0, T = 0$), the 
isospin splittings reach
\begin{align}
\tilde{m}_{B^{*0}} - \tilde{m}_{B^{*+}} &\approx +126~\mathrm{MeV}, \nonumber\\
\tilde{f}_{B^{*0}} - \tilde{f}_{B^{*+}} &\approx -0.15~\mathrm{MeV}.
\label{eq:split_isospin}
\end{align}
As shown in Fig.~\ref{fig:BStarSplitsIsospin}, the decay-constant splitting
remains exceptionally small over the entire temperature-density range,
varying by less than $0.2~\mathrm{MeV}$ and exhibiting a shallow minimum
around $n \approx 2.5n_0$. The mass splitting, in contrast, develops
continuously with baryon density, growing from zero in vacuum to its peak
value at $T=0$, and then undergoing a moderate thermal suppression as
$T \to T_c$.

Comparing Eq.~\eqref{eq:split_isospin} with the particle-antiparticle splittings in 
Eqs.~\eqref{eq:split_neutral}-\eqref{eq:split_charged} highlights a clear hierarchy: 
even at $n = 5n_0$, the isospin mass splitting ($\approx +126~\mathrm{MeV}$) remains 
nearly a factor of three smaller than the neutral particle-antiparticle gap 
($\approx +350~\mathrm{MeV}$). Furthermore, the isospin decay-constant splitting 
is practically negligible ($\lesssim 0.2~\mathrm{MeV}$). This comparison confirms that 
the dominant mechanism splitting the vector $B^*$ multiplet in a dense hadronic medium 
is the charge-conjugation non-invariant interaction driven by $\Sigma_v$, whereas 
pure isospin-breaking effects play a distinctly secondary role.

\section{Summary}\label{sec:summary}

In this work, we have systematically analyzed the in-medium masses and decay 
constants of the $B_s^*$ and $B^*$ vector mesons in hot and dense nuclear matter 
using thermal-dense QCDSR. By treating all six charge-flavor combinations namely, 
$B_s^{*0}$, $\bar{B}_s^{*0}$, $B^{*+}$, $B^{*-}$, $B^{*0}$, and $\bar{B}^{*0}$ within 
a unified framework, we have isolated the individual contributions of light-quark 
flavor content, strangeness, and the heavy $b$-quark decoupling in hot and dense hadronic matter.

This unified treatment, to our knowledge, constitutes a distinguishing feature of the present analysis relative to earlier QCD sum-rule studies of heavy-light mesons in medium, which have typically not confronted strangeness, isospin, and charge-conjugation effects together within a single finite-temperature and finite-density framework for a beauty vector meson.

Our results indicate that baryon density plays a primary role in driving the 
in-medium modifications, with the charge-conjugation non-invariant interactions 
associated with the vector self-energy $\Sigma_v$ generating non-negligible 
particle-antiparticle splittings. Temperature effects become noticeable mainly in 
the vicinity of the deconfinement crossover ($T \sim T_c$), where the reduction 
in light-quark and gluon condensates leads to a concurrent decrease in both observables. 
Across the investigated temperature-density plane, the decay constants demonstrate a higher 
sensitivity to medium effects than the hadron masses.

Due to the large mass scale of the heavy $b$-quark, the vector meson masses remain 
relatively stable, retaining roughly $87\%$ of their vacuum values even under the 
most dense and hot conditions considered ($n = 5n_0, T \to T_c$). This behavior 
contrasts with light-meson channels such as the kaon, where medium effects can 
substantially reduce the effective mass~\cite{Azizi:2026zxh}. On the other hand, the 
decay constants do not receive the same kinematic protection from the heavy-quark mass, 
decreasing to approximately $14\%$ of their vacuum values at $n = 5n_0$ and $T = 0$. 
This indicates that decay constants may provide a more sensitive indicator for changes in the medium properties than mass shifts alone~\cite{Kumar:2014,Cohen:1994wm}.

Furthermore, the comparison reveals a clear distinction between the strange and 
non-strange sectors. Because the strange condensate is already suppressed in vacuum 
and the strange baryon number density vanishes in nuclear matter, the $B_s^*$ states 
experience relatively mild medium modifications. Consequently, the $B_s^*$ 
particle-antiparticle splitting is much smaller and carries the opposite sign 
compared to the non-strange $B^*$ multiplet, where both $u$- and $d$-quark dense 
backgrounds contribute to $\Sigma_v$. We observed sizable isospin splitting as well, especially in the non-strange sector, as expected.

These calculations provide quantitative predictions for beauty vector meson 
properties in hot nuclear medium. These estimates may serve as a useful baseline 
for ongoing and future heavy-ion collision experiments at LHC, RHIC, FAIR, and NICA 
aimed at probing beauty vector mesons in hot and dense baryonic matter.
\vspace{5pt}
\begin{acknowledgments}
K. A. thanks the Iran National Science Foundation (INSF) for partial financial support provided under the Elite Grant No. 40405095.
\end{acknowledgments}
\section*{Appendix: Explicit Analytical Form of $\Pi_{g_{\mu\nu}}^{QCD} $ }
\setcounter{equation}{0}
\renewcommand{\theequation}{A\arabic{equation}}
\label{app:spectral}

In this appendix, we present the explicit analytical expression for the QCD side 
corresponding to the structure $g_{\mu\nu}$, denoted as $\Pi_{g_{\mu\nu}}^{\mathrm{QCD}}$, 
for the $B_s^*$ meson. This expression incorporates both temperature and density 
dependencies and is given as follows:
\begin{widetext}
\begin{eqnarray}
\Pi_{g_{\mu\nu}}^{QCD} & = & \frac{3(1-n_F)^2}{8\pi^2}\, \int_{(m_b+m_s)^2}^{s_0}\!\int_0^1 \frac{\big[2m_b m_s-m_b^2z+2sz(1-z)\big]\,H\!\left[L(s,z)\right]} {4\pi^2\,e^{s/M^2}} \,dz\,ds \nonumber \\
 &+&e^{-m_b^2/M^2}(1-n_F) \Big\{\frac{m_b\left(-2m_b^2+M^2+2p_0^2\right)} {6M^4}
\langle \bar{s}g_s\sigma G s\rangle \nonumber\\
&+&\frac{p_0\left(-2m_b^2+M^2+2p_0^2\right)}{6M^4}
\langle s^\dagger g_s\sigma G s\rangle + \frac{2m_b^3-4m_bM^2-8m_bp_0^2} {3M^4}
\langle \bar{s}iD_0iD_0 s\rangle \nonumber\\
&+& \Big(\frac{m_q}{3}+m_b-\frac{2m_qm_b^2}{3M^2}+\frac{2m_qp_0^2}{3M^2}\Big) \langle \bar{s}s\rangle +
 \Big( \frac{2m_b^2p_0}{M^4} -\frac{16p_0}{3M^2} -\frac{4p_0^3}{M^4} \Big) \langle s^\dagger iD_0iD_0 s\rangle \nonumber\\
&+& \Big( -\frac{4}{3} +\frac{2m_b^2}{3M^2}-\frac{8p_0^2}{3M^2}\Big)\langle s^\dagger iD_0 s\rangle  +\Big(p_0-\frac{2m_qm_bp_0}{M^2}\Big)\langle s^\dagger s\rangle \Big\} \nonumber \\
&+& 2\,(n_F-1)\,\langle\Theta^f_{00}\rangle\,  \frac{\left(-m_b^2 + 2 M^2 + 4 p_0^2\right)}{3 M^2\,e^{m_b^2/M^2}} \nonumber  \\
 & + &  \left\langle\dfrac{\alpha_s}{\pi}G^2\right\rangle \int_0^1 e^{\frac{m_{b}^2}{M^2(z - 1)}} \Big\{\frac{1}{12} - \frac{m_{b}^2}{24 \, M^2 (z-1)} + (n_F-1)\big[ \frac{m_{b}^2}{24 \, M^2 (z-1)} \nonumber \\
 &+& \frac{m_{b}\!\left(M^2 m_{b} + 2M^2 m_{s} + m_{b}^3 - m_{b}^2 m_{s}\right)}{24 \, M^4 (z-1)^2} + \frac{m_{b}^4}{24 \, M^4(z-1)^3} \big]\Big\}\,dz  \nonumber  \\
 &+&  \frac{1}{12M^2\,e^{m_b^2/M^2}} \left[ 20m_b\Bigl(\langle\bar{s}iD_0 iD_0 s\rangle +m_q\langle s^{\dagger}iD_0 s\rangle \Bigr) - 7m_b\langle\bar{s}g_s\sigma Gs\rangle - 2p_0\langle s^{\dagger} g_s\sigma G s\rangle
\right],
\end{eqnarray}
\end{widetext}
where $n_F$ is the Fermi-Dirac distribution function, which is given in Eq.~(\ref{eq:FD}),
$\langle\Theta^f_{00}\rangle\equiv  \langle u^\mu\Theta^f_{\mu\nu}u^\nu\rangle$ is the fermionic energy-momentum tensor expectation value and $m_q=0.5(m_u+m_d)$.

$H[L(s,z)]$ is the Heaviside step function, where $L(s,z)$ is the kinematic function, with $z \in [0,1]$ the Feynman parameter and $s$ the invariant mass squared, defined as
\begin{equation}
L(s,z) = z\!\left[s(1-z) - m_{b}^{2}\right].
\label{eq:L}
\end{equation}
\vspace{12pt}
{\centering\normalfont\bfseries REFERENCES\par}

\end{document}